\documentclass{jfm}

\usepackage{graphicx}
\usepackage{natbib}

\ifCUPmtlplainloaded \else
  \checkfont{eurm10}
  \iffontfound
    \IfFileExists{upmath.sty}
      {\typeout{^^JFound AMS Euler Roman fonts on the system,
                   using the 'upmath' package.^^J}%
       \usepackage{upmath}}
      {\typeout{^^JFound AMS Euler Roman fonts on the system, but you
                   dont seem to have the}%
       \typeout{'upmath' package installed. JFM.cls can take advantage
                 of these fonts,^^Jif you use 'upmath' package.^^J}%
      }
  \else
  \fi
\fi

\ifCUPmtlplainloaded \else
  \checkfont{msam10}
  \iffontfound
    \IfFileExists{amssymb.sty}
      {\typeout{^^JFound AMS Symbol fonts on the system, using the
                'amssymb' package.^^J}%
       \usepackage{amssymb}%
         \let\leq=\leqslant
         \let\geq=\geqslant
      }{}
  \fi
\fi

\ifCUPmtlplainloaded \else
  \IfFileExists{amsbsy.sty}
    {\typeout{^^JFound the 'amsbsy' package on the system, using it.^^J}%
     \usepackage{amsbsy}}
    {\providecommand\boldsymbol[1]{\mbox{\boldmath $##1$}}}
\fi

\providecommand\bnabla{\boldsymbol{\nabla}}
\providecommand\bcdot{\boldsymbol{\cdot}}

\newcommand\Real{\mbox{Re}} 
\newcommand\Imag{\mbox{Im}} 
\newcommand\Rey{\mbox{\textit{Re}}}  

\newsavebox{\astrutbox}
\sbox{\astrutbox}{\rule[-5pt]{0pt}{20pt}}

\newcommand\etc{etc.\ }
\newcommand\eg{e.g.\ }

\usepackage{color}
\usepackage{psfrag}
\usepackage{enumerate}
\usepackage[percent]{overpic}
\usepackage[normalem]{ulem}

\def\aa{\mathbf{a}}

\def\bb{\mathbf{b}}

\def\bdelta{\boldsymbol{\delta}}

\def\ex{\mathbf{e}_x}

\def\nn{\mathbf{n}}
\def\tt{\mathbf{t}}
\def\xx{\mathbf{x}}

\def\FF{\mathbf{F}}

\def\00{\mathbf{0}}

\def\U{U} 					

\def\UU{\mathbf{U}}
\def\QQ{\mathbf{Q}}

\def\uu{\mathbf{u}}			
\def\qq{\mathbf{q}}

\def\pa{p^\dag}
\def\uua{\mathbf{u}^\dag}	
\def\qqa{\mathbf{q}^\dag}

\def\Ua{U^\dag} 					
\def\Va{V^\dag} 
\def\Pa{P^\dag} 

\def\UUa{\mathbf{U}^\dag}
\def\QQa{\mathbf{Q}^\dag}

\def\nnu{\Rey^{-1}}

\def\ev{\sigma}

\def\s2{G^2}
\def\DUs2{\nabla_\UU \s2}
\def\DQs2{\nabla_\QQ \s2}
\def\DQFs2{\nabla_{\QQ_\FF} \s2}
\def\DFs2{\nabla_\FF \s2} 
\def\DQws2{\nabla_{\QQ_w} \s2}
\def\DUws2{\nabla_{\UU_w} \s2}

\def\rl{{l_c}} 

\def\grow{{\sigma_r}} 
\def\puls{{\sigma_i}}

\newcommand\be{\begin{equation}}
\newcommand\ee{\end{equation}}

\definecolor{mygreen}{rgb}{0,0.5,0}

\title[A variational approach to recirculation length reduction]
{Controlled reattachment in separated flows: 
a variational approach to  recirculation 
length reduction}

\author[E. Boujo and F. Gallaire]%
{E. Boujo$^1$%
  \thanks{Email address for correspondence: edouard.boujo@epfl.ch}
and F. Gallaire$^1$}

\affiliation{$^1$LFMI, 
\'Ecole Polytechnique F\'ed\'erale de Lausanne,
CH-1015 Lausanne, Switzerland}

\pubyear{}
\volume{}
\pagerange{}
\date{?; revised ?; accepted ?. - To be entered by editorial office}

\begin{document}

\maketitle

\begin{abstract}
A variational technique is used to derive analytical expressions for the sensitivity of recirculation length to steady forcing in separated flows.
Linear sensitivity analysis is applied to the two-dimensional steady flow past a circular cylinder for Reynolds numbers $40 \leq \Rey \leq 120$, both in the subcritical and supercritical regimes. Regions which are the most sensitive to volume forcing and wall blowing/suction are identified.
Control configurations which reduce the recirculation length are designed based on the sensitivity information, in particular  small cylinders used as control devices in the wake of the main cylinder, and fluid suction at the cylinder wall.
Validation against full non-linear Navier-Stokes calculations shows excellent agreement for small-amplitude control.
The linear stability properties of the controlled flow are systematically investigated. At moderate Reynolds numbers, we observe that regions where control reduces the recirculation length correspond to regions where it has a stabilising effect on the most unstable global mode associated to vortex shedding, 
while this property does not hold any more at larger Reynolds numbers.
\end{abstract}



\section{Introduction}

Separation occurs in  flow configurations with abrupt geometry changes or strong  adverse pressure gradients.
In practical engineering applications, separation is generally associated with low-frequency fluctuations which can have undesirable effects, \eg  deterioration of vehicle performance,  fatigue of mechanical structures, \etc  
The control of separated flows is therefore an active research area.
Part of the ongoing research work focuses on the laminar regime, 
where separated flows are steady at low Reynolds number and become unsteady above a threshold value.
In this regime, stability theory can help design control strategies by providing insight into the physical phenomena involved in the transition to unsteadiness through, for example, 
the transient growth of particular initial perturbations, or the bifurcation of unsteady eigenmodes.
Examples of separated flows commonly studied as archetypical configurations because of their fundamental interest
include bluff bodies (\eg square and circular cylinders), backward-facing steps, bumps, stenotic geometries, and pressure-induced separations over flat plates.

Sensitivity analysis  uses a variational approach to calculate  efficiently the linear sensitivity of some quantity to a modification of the flow or to a given actuation, thus suppressing the need to resort to exhaustive parametric studies. 
\citet{Hill92AIAA} applied sensitivity analysis to the flow past a cylinder and computed the sensitivity of the most unstable growth rate to passive control by means of a second smaller cylinder, and successfully reproduced most sensitive regions previously identified  experimentally by \citet{Stry90}.
Since then, sensitivity analysis gained popularity and was applied to evaluate sensitivity 
to flow modification or to passive control in various flows, both in local and global frameworks.
For example, \citet{cor01tcfd} designed a control strategy based on such a variational technique in order to reduce optimal transient growth in boundary layers. This was achieved by computing the sensitivity of an objective function involving energy, which was then iteratively minimized. 
Such quadratic cost functionals are very often employed in control theory, but sensitivity analysis can be applied to non-quadratic quantities as well.
\citet{Bew01} minimized several kinds of cost functionals and successfully relaminarized the turbulent flow in a plane channel using wall transpiration.
\citet{Bot03} used a variational approach to compute the sensitivity of eigenvalues to base flow modification in the parallel plane Couette flow, as  well as the most destabilising modification.
\citet{mar08cyl} studied the sensitivity of the cylinder flow leading eigenvalue to base flow modification and to steady forcing in the bulk and, again, reproduced the regions of \citet{Stry90}.
 \citet{Mel10} managed to control the first oscillating eigenmode in the compressible flow past a slender axisymmetric body by considering its sensitivity to  steady forcing, both in the bulk (with mass, momentum or energy sources) and at the wall (with blowing/suction or heating).
Recently, \citet{Bra11} also applied sensitivity analysis to evaluate the effect of steady control on noise amplification (maximal energy amplification under harmonic forcing in steady-state regime) in a globally stable flat-plate boundary layer.

In the present study, sensitivity analysis is applied to another quantity of interest in separated flows: the length of the recirculation region $\rl$. 
Many authors observed 
that in  separated flows the recirculation length increases with $\Rey$ (below the onset of instability): 
circular cylinder 
(experimental study by  \citet{Tan56}, 
numerical study by
 \citet{Gia07}), 
backward-facing step 
(experimental study by  \citet{Acr68},
numerical study by \citet{bar02}), 
wall-mounted bump (numerical study by  \citet{Mar03},
experimental study by  \citet{pas12}), 
\etc
As the recirculation region gets longer, both maximal backward flow and maximal shear increase. 
From a local stability viewpoint, this tends to destabilise the flow.
In addition, since the shear layer elongates,
 incoming or developing perturbations are amplified over a longer distance while advected downstream,
and any region of absolute instability is increased in length too.
When the flow becomes unstable and unsteady, as is the case for the cylinder flow above threshold ($\Rey > \Rey_c$), the mean recirculation length decreases \citep{nis78}. 
This is interpreted as the result of a mean flow correction, and the decrease in the mean value of $\rl$ naturally appears as a characteristic global order parameter of the bifurcation \citep{zie97}.

The recirculation length therefore appears as a relevant macroscopic scalar parameter to characterize separated flows.
This motivates the design of control strategies which directly target $\rl$, rather than eigenmode growth rates, transient growth, or noise amplification.
In other words, we propose control strategies which do not focus on the fate of perturbations but  act upon a feature of the base flow itself.

We choose to design control configurations based on the steady-state base flow, and consider both subcritical and supercritical Reynolds numbers, $40 \leq \Rey \leq 120$.
In the supercritical regime $\Rey > \Rey_c$, the uncontrolled steady-state base flow is linearly unstable and the actual flow observed in experiments or numerical simulations is unsteady; 
but the sensitivity of the steady-state recirculation length is of interest since reducing $\rl$ might restabilise the flow.
The stability of the controlled flow will be assessed systematically to determine when this approach is relevant.

This paper is organized as follows.
Section~\ref{sec:problem} details the problem formulation and numerical methods. In particular, analytical expressions are derived for sensitivity of  recirculation length to base flow modifications and to steady control, both volume forcing and wall blowing/suction.
Results are presented in section~\ref{sec:results}: regions sensitive to forcing are identified, and several control configurations which allow to reduce $\rl$ are selected to illustrate the method and to validate the sensitivity analysis against fully non-linear simulations. 
The linear stability properties of these controlled flows are investigated and discussed in section~\ref{sec:stability}.
Conclusions are drawn in section~\ref{sec:conclusion}.

\section{Problem formulation and numerical methods}
\label{sec:problem}

\begin{figure}
\vspace{0.4 cm}
\centerline{
	\begin{overpic}[width=8 cm,tics=10]{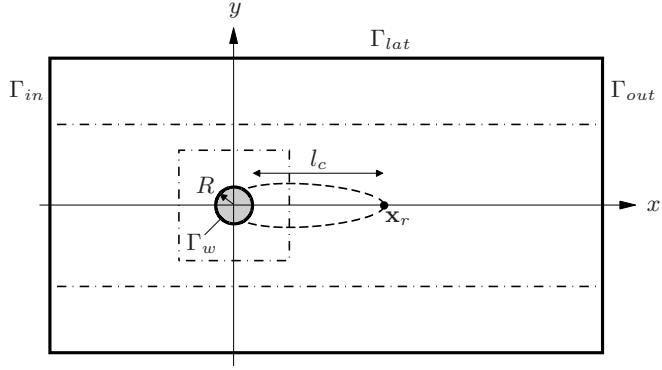}
    \put(-5, 45) {$\Gamma_{in}$}
    \put(95, 45) {$\Gamma_{out}$}
    \put(55, 53.5) {$\Gamma_{lat}$}
    \put(24.5, 19.5) {$\Gamma_{w}$}
    \put(101, 26) {$x$}
    \put(31.5, 59) {$y$}
    \put(57.5, 24.2) {$\xx_r$} 
    \put(45, 33.5) {$\rl$}
    \put(26, 29) {$R$}
  	\end{overpic}
}
  \caption{
  Schematic of the problem geometry and computational domain.}
\label{fig:geom_and_domain}
\end{figure}

The sensitivity of the recirculation length in a two-dimensional  incompressible cylinder flow is investigated.
A cylinder of radius $R$ is located in a uniform flow. 
The fluid motion is described by the velocity field $\UU=(U,V)^T$ of components $U$ and $V$ in the streamwise and cross-stream directions $x$ and $y$, and the pressure field $P$.
The state vector $\QQ=(\UU,P)^T$ is solution of the two-dimensional incompressible Navier--Stokes equations 
\be \label{eq:NS}
	\bnabla \bcdot \UU = 0,
	\quad
	\partial_t \UU +  \bnabla \UU \bcdot  \UU + \bnabla  P - \nnu \bnabla^2  \UU = \FF
\ee
where $\Rey=U_\infty D/\nu$ is the Reynolds number based on the cylinder diameter $D=2R$, the freestream velocity $U_\infty$ and the fluid kinematic viscosity $\nu$, and $\FF$ is a steady volume forcing in the bulk.
The following boundary conditions are prescribed:
uniform velocity profile $\UU_b=(\U_\infty,0)^T$ at the inlet $\Gamma_{in}$,
symmetry condition $\partial_y  U_b=0, V_b=0$ on lateral boundaries $\Gamma_{lat}$, 
outflow condition $-P_b \nn+\nnu\bnabla \UU_b\bcdot\nn=\00$ on $\Gamma_{out}$,
where $\nn$ is the normal unit vector oriented outward the domain,
blowing/suction $\UU_b=\UU_w$ on the wall control region $\Gamma_c$, 
and  no-slip condition $\UU_b=\00$ on the remaining cylinder wall region $\Gamma_w \setminus \Gamma_c$.
In this paper attention is restricted to steady flows $\QQ_b(x,y)$ which satisfy:
\be \label{eq:NSsteady}
	\bnabla \bcdot \UU_b = 0,
	\quad
	\bnabla \UU_b \bcdot \UU_b + \bnabla  P_b - \nnu \bnabla^2 \UU_b = \FF.
\ee

\subsection{Sensitivity of recirculation length}
Assuming the flow is symmetric  with respect to the symmetry axis $y=0$, the recirculation length is defined as the distance from the cylinder wall rearmost point $(R,0)$  to the reattachment point $\xx_r = (x_r,0)$ 
as shown in figure~\ref{fig:geom_and_domain}:
\be
	\rl = x_r-R.
	\label{eq:rl1}
\ee
The reattachment point is characterized by zero streamwise velocity, $U(x_r,0)=0$, and can therefore be computed with a bisection method on  $U_c(x)=U(x,0)$ along the symmetry axis.

Throughout this study, only flow modifications and forcings which are symmetric with respect to the symmetry axis will be considered; they result in symmetric flows, thus ensuring that the recirculation length (\ref{eq:rl1}) is well defined.

\subsubsection{Sensitivity to base flow modification}

Considering a small  modification of the base flow 
$\bdelta \QQ$,
the variation of the recirculation length 
is expressed at first order as
\be 
	\delta \rl = (\bnabla_\QQ \rl \,|\, \bdelta  \QQ)
	\label{eq:dl}
\ee
where $\bnabla_\QQ \rl = (\bnabla_\UU \rl , \bnabla_P \rl)^T$ is the sensitivity to base flow modification,
$ (\aa \,|\, \bb) = \int_{\Omega} \bar\aa \bcdot \bb \,\mathrm{d}\Omega$ denotes the two-dimensional inner product  for real or complex fields, 
and the overbar  stands for complex conjugate.

\begin{figure}
  \psfrag{Uc}[r][][1][-90]{$U_c$}
  \psfrag{H}[r][][1][-90]{$G$}
  \psfrag{x}[][][1][0]{$x$}
  \centerline{
  \begin{overpic}[height=4.3 cm,tics=10]{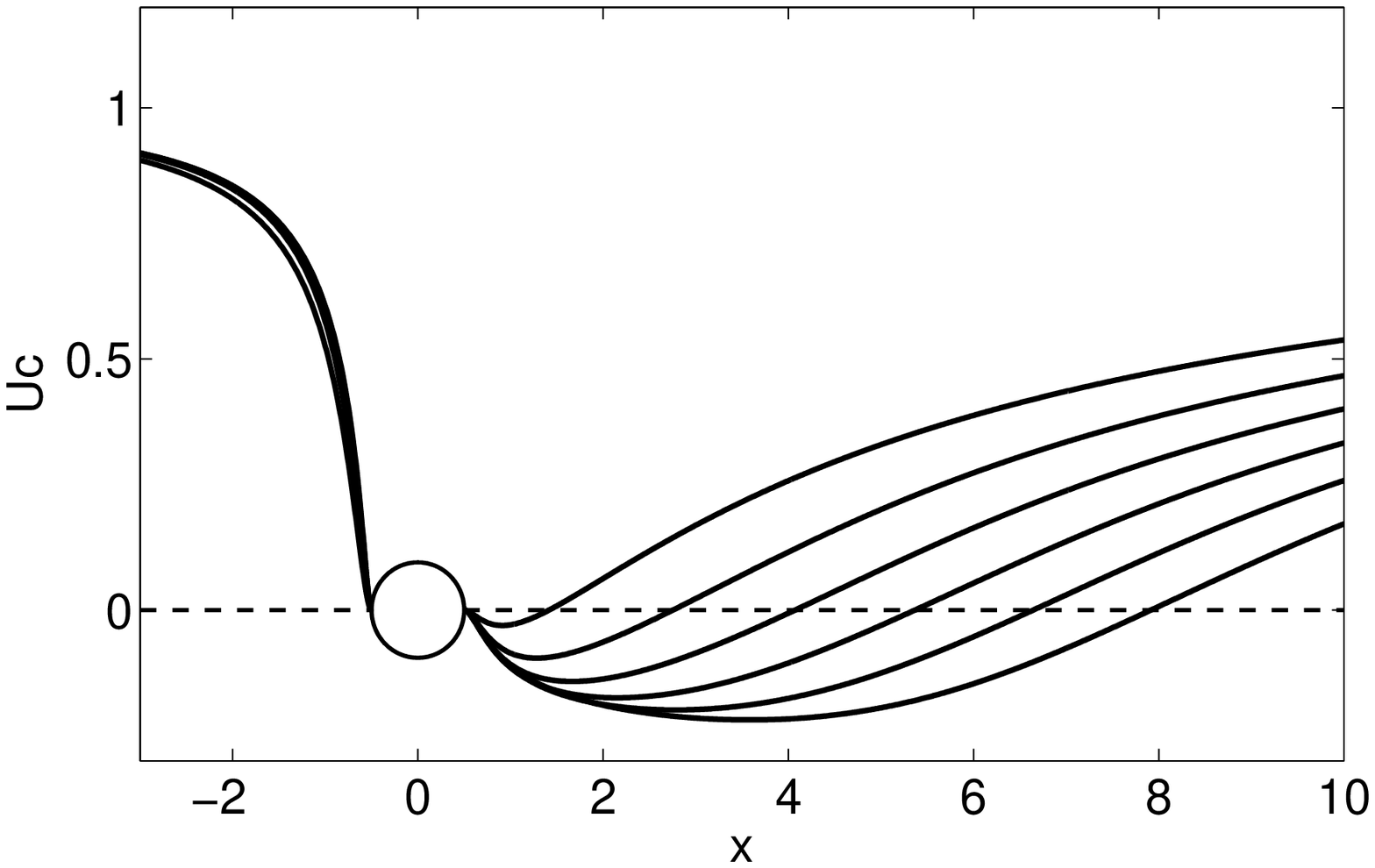}
     \put(12.5,57.5) {$(a)$} 
     \put(67,38){\footnotesize $\Rey=20$} 
     \put(76,12) {\footnotesize $\Rey=120$} 
  \end{overpic}
  \hspace{0.1 cm}  
  \begin{overpic}[height=4.3 cm,tics=10]{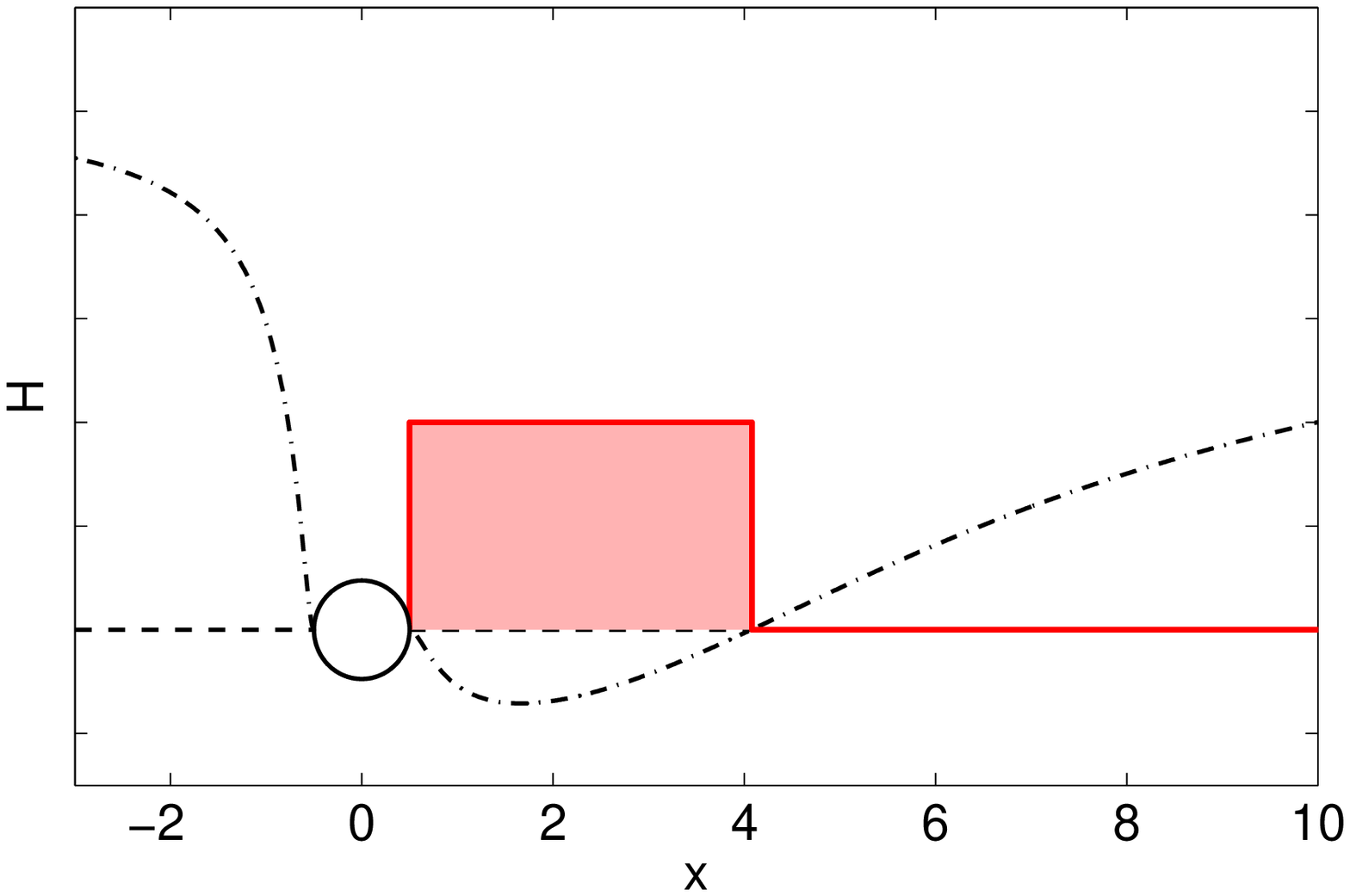}
     \put(8,60) {$(b)$} 
     \put(37,37) {$G=1$} 
     \put(74,22) {$G=0$} 
     \put(54.5,16) {$x_r$} 
  \end{overpic}
  }
  \caption{ 
 $(a)$ Streamwise velocity along the symmetry axis 
 at $\Rey=20, 40 \ldots 120$.
 $(b)$~Integrand in the definition of  the recirculation length (\ref{eq:rl2b}), illustrated at $\Rey=60$.
 }
\label{fig:profiles}
\end{figure}

To allow for the calculation of this sensitivity, the recirculation length is rewritten as
\be
	\rl 
	=  \int_{R}^{\infty} H \left( - U_c(x) \right) \,\mathrm{d}x
	=  \int_{R}^{\infty} G \left( x        \right) \,\mathrm{d}x,
	\label{eq:rl2b}
\ee
where $U_c(x)=U(x,0)$ is the streamwise velocity on the symmetry axis 
and $H$ is the Heaviside step function defined as 
$H(\alpha)=0$ for $\alpha<0$ and 
$H(\alpha)=1$ for $\alpha>0$.
As illustrated in figure \ref{fig:profiles}, the integrand 
is equal to 1 in the recirculation region where $U_c(x)<0$, and is equal to 0 downstream, 
therefore integrating along $x$ from the rear stagnation point gives the recirculation length.

Using the same Lagrangian formalism as  \citet{Hill92AIAA}, the recirculation length variation due to a base flow modification 
$\bdelta \QQ =(\bdelta\UU, \delta P)^T$ is obtained as:
\begin{subeqnarray} \label{eq:dlc}
 	\delta  \rl
& = & \lim_{\epsilon \rightarrow 0} \frac{1}{\epsilon} \left[ \rl(\QQ+\epsilon \bdelta \QQ)-\rl(\QQ) \right]
\slabel{eq:dlc0}
 \\
& = & \lim_{\epsilon \rightarrow 0} 
\frac{1}{\epsilon} \int_{R}^{\infty} \left[ H \left( - U_c(x)-\epsilon \delta U_c(x) \right) -  H \left( -U_c(x) \right) \right] \,\mathrm{d}x
\slabel{eq:dlc1}
 \\
& = & 
\int_{R}^{\infty}  -\left.\frac{ \mathrm{d}H }{ \mathrm{d}U }\right|_{U=-U_c(x)} \,\delta U_c(x) \,\mathrm{d}x
 \slabel{eq:dlc2}
 \\
& = & 
\int_{R}^{\infty}  \left(  \frac{\mathrm{d} U_c}{\mathrm{d} x}(x) \right)^{-1} 
\frac{ \mathrm{d}G }{ \mathrm{d}x }(x) \,\delta U_c(x) \,\mathrm{d}x
\slabel{eq:dlc3} 
\\
& = & 
\int_{R}^{\infty} -  \left(  \frac{\mathrm{d} U_c}{\mathrm{d} x}(x) \right)^{-1}  \delta(x-x_r) \,\delta U_c(x) \,\mathrm{d}x
\slabel{eq:dlc4} 
\\
& = & - \frac{ \delta U_c(x_r) } { \mathrm{d}_x U_c|_{x=x_r}},
\slabel{eq:dlc5} 
\end{subeqnarray}
where (\ref{eq:dlc3}) comes from the differentiation of $G(x)=H(-U_c(x))$ using the chain rule,
$\displaystyle \frac{\mathrm{d} G}{\mathrm{d} x}(x) $
$\displaystyle= -\left.\frac{\mathrm{d} H}{\mathrm{d} U}\right|_{U=-U_c(x)} \frac{\mathrm{d} U_c}{\mathrm{d} x}(x)$,
and (\ref{eq:dlc4}) is the result of 
$\displaystyle \frac{\mathrm{d} G}{\mathrm{d} x}(x)=-\delta(x-x_r)$ with $\delta(x)$ the Dirac delta function, since $G$ jumps from 1 to 0 at $x=x_r$.
If, for example, the streamwise velocity at the original reattachment point increases, then the recirculation region is shortened: this is understood physically as $U_c(x_r)$ becoming positive and  the reattachment point moving upstream, 
while mathematically  $\delta U_c(x_r)>0$ and 
 (\ref{eq:dlc}) yield $\delta \rl <0$ (because $\mathrm{d}_x U_c|_{x=x_r}>0$, see figure \ref{fig:profiles}$(a)$).

The sensitivity to base flow modification is  identified  as:
\be 
	\bnabla_\UU \rl = -\frac{1}{\mathrm{d}_x U_c|_{x=x_r}}  
	\left( 	\begin{array}{cc}
		 	\delta(x_r,0) \\ 0
			\end{array} 
	\right),
	\quad
	\nabla_P \rl = 0,
	\label{eq:sens_open}
\ee
where $\delta(x,y)$ is the two-dimensional Dirac delta function, such that the inner product (\ref{eq:dl}) between the two fields $\bnabla_\QQ \rl$ and  $\bdelta \QQ$ is indeed $\delta \rl$ as expressed by (\ref{eq:dlc}).

In wall-bounded flows, where reattachment occurs at the wall (for example behind a backward-facing step or a bump, or on a flat plate with adverse pressure gradient), the reattachment point is not characterized by zero streamwise velocity $U(x_r,0)=0$,  but instead by zero wall shear stress, i.e. 
$\partial_y U|_{x=x_r,y=0}=0$ 
with the wall assumed horizontal and located at $y=0$ for the sake of simplicity. In this case  the sensitivity reads:
\be 
	\bnabla_\UU \rl = -\frac{1}{\partial_{xy} U|_{x=x_r,y=0}}  
	\left(	\begin{array}{cc}
			\delta(x_r,0) \partial_y \\ 0
			\end{array} 
	\right),
	\quad
	\nabla_P \rl = 0.
	\label{eq:sens_wallbounded}
\ee

The sensitivity field $\bnabla_\QQ \rl$ in (\ref{eq:sens_open})-(\ref{eq:sens_wallbounded}) is valid for any arbitrary base flow modification $\bdelta\UU$.
As noted by \citet{Bra11}, it is possible to derive a restricted sensitivity field for divergence-free base flow modifications.
In the case of the cylinder flow, where (\ref{eq:sens_open}) results in a localized Dirac delta function at the reattachment point in the $x$ direction only, 
this restricted sensitivity field appears to present a dipolar structure.

\subsubsection{Sensitivity  to forcing}

Now the sensitivity of the recirculation length to steady forcing is investigated.
One considers a small-amplitude forcing: volume force in the bulk $\bdelta \FF(x,y)$,
or blowing/suction $\bdelta \UU_w$ on part of the cylinder wall. The recirculation length variation at first order is 
\be 
	\delta \rl = 
	(\bnabla_\FF \rl \,|\, \bdelta \FF) + 
	\langle \bnabla_{\UU_w} \rl \,|\, \bdelta \UU_w \rangle
	\label{eq:RecircLengthVarForce}
\ee
where $\langle \aa \,|\, \bb \rangle = \int_{\Gamma_c} \bar\aa \bcdot \bb \,\mathrm{d}\Gamma$ denotes  the one-dimensional inner product on the control boundary.
The same Lagrangian formalism as in the previous section yields the sensitivities
\be 
 	\bnabla_\FF \rl = \UUa,
 	\quad
	\bnabla_{\UU_w} \rl = \UUa_w = -\Pa \nn - \nnu \bnabla \UUa \bcdot \nn,
\ee
where the so-called adjoint base flow $\QQa=(\UUa,\Pa)^T$ is solution of the linear, non-homogeneous system of equations
\be 
	\bnabla \bcdot \UUa  =0,
 	\quad
  	-\bnabla \UUa \bcdot  \UU_b + \bnabla \UU_b^T \bcdot  \UUa
	- \bnabla \Pa - \nnu \bnabla^2 \UUa = \bnabla_\UU \rl,  
\label{eq:adjBF-lc}
\ee
with the boundary conditions 
$\UUa=\00$ on $\Gamma_{in} \cup \Gamma_w \cup \Gamma_c$,
symmetry condition $\partial_y  \Ua=0, \Va=0$ on $\Gamma_{lat}$, 
and $\Pa\nn+\nnu\bnabla\UUa\bcdot\nn+\UUa(\UU_b\bcdot\nn)=\00$ on $\Gamma_{out}$.

\subsubsection{Effect of a small control cylinder}
\label{sec:smallctrlcyl}

It is of practical interest to study the effect of a particular kind of passive control on recirculation length and eigenvalues, namely a small control cylinder of diameter $d \ll D$ similar to the one used by \citet{Stry90} to suppress vortex shedding in a limited range of Reynolds number above instability threshold.
%
The effect on the base flow of a small control cylinder located at $\xx_c=(x_c,y_c)$ is modelled as a steady volume force of same amplitude as the drag force acting on this control cylinder, and of opposite direction:
\be
	\bdelta \FF(x,y) = -\frac{1}{2} d C_d(x,y) ||\UU_b(x,y)|| \UU_b(x,y) \delta(x-x_c,y-y_c) 
	\label{eq:SmallCylForce}
\ee
where $C_d$ is the drag coefficient of the control cylinder and depends on the local Reynolds number $\Rey_d(x,y)=||\UU_b(x,y)|| d/\nu$.
Finally, variations of the recirculation length and eigenvalue (see section \ref{sec:stab}) are calculated from
$\delta \rl = (\bnabla_\FF \rl \,|\, \bdelta \FF)$ and 
$\delta \ev = (\bnabla_\FF \ev \,|\, \bdelta \FF)$:
\begin{eqnarray}
\delta \rl(x_c,y_c) &=& -\frac{1}{2} d C_d(x_c,y_c) ||\UU_b(x_c,y_c)|| \bnabla_\FF \rl(x_c,y_c) \bcdot \UU_b(x_c,y_c),
	\label{eq:SmallCylEffectRL}
\\
	\delta \ev(x_c,y_c) &=& -\frac{1}{2} d C_d(x_c,y_c) ||\UU_b(x_c,y_c)|| \bnabla_\FF \ev (x_c,y_c) \bcdot \UU_b(x_c,y_c).
	\label{eq:SmallCylEffectEV}
\end{eqnarray}

For a diameter $d=D/10$ and for the set of Reynolds numbers $\Rey$ and locations $(x_c,y_c)$ chosen hereafter, the Reynolds number of the control cylinder falls in the range $1 \leq \Rey_d \leq 15$.
The expression of \cite{Hill92AIAA} has been generalized in this range according to 
 $C_d(\Rey_d) = a + b \Rey_d^c$,
 based on a set of 
experimental data from \citet{fin53} and \citet{tri59}
and from numerical results obtained by the authors, yielding
$a=0.8558, b= 10.05, c=-0.7004$.
%

\subsection{Linear stability}
\label{sec:stab}

Writing the flow as the superposition of a steady base flow and time-dependent small perturbations, 
$\QQ(x,y,t) = \QQ_b(x,y) + \qq'(x,y,t)$,
linearising the Navier--Stokes equations~(\ref{eq:NS}) and using the normal mode expansion
$\qq'(x,y,t) = \qq(x,y) e^{\ev t}$, with $\ev = \grow + i \puls$, the following system of equations is obtained:
\be \label{eq:evp}
	\bnabla \bcdot \uu = 0, 
	\quad
	\ev \uu + \bnabla \uu \bcdot \UU_b + \bnabla \UU_b \bcdot \uu + \bnabla p - \nnu \bnabla^2 \uu = \00,
\ee
together with the following boundary conditions:
$\uu=\00$ on $\Gamma_{in} \cup \Gamma_w \cup \Gamma_c$,
symmetry condition $\partial_y u=0, v=0$ on $\Gamma_{lat}$, 
and  outflow condition $- p \nn+\nnu\bnabla \uu \bcdot \nn=\00$ on $\Gamma_{out}$.
Solving this generalized eigenvalue problem yields global modes $\qq$ and associated growth rate $\grow$ and pulsation $\puls$.

The sensitivity of an eigenvalue to base flow modification, defined by 
$\delta \ev = (\bnabla_\UU \ev \,|\, \bdelta \UU)$, 
can be computed as 
\be 
	\bnabla_{\UU} \ev = -\bnabla \bar\uu^T \bcdot \uua + \bnabla \uua \bcdot \bar\uu,
\ee
where $\qqa=(\uua,\pa)^T$ is the adjoint mode associated with $\ev$.
The sensitivity to steady forcing, 
defined by 
$\delta \ev = (\bnabla_\FF \ev \,|\, \bdelta \FF) + \langle \bnabla_{\UU_w} \ev \,|\, \bdelta \UU_w \rangle$, 
can be computed as
\be 
 	\bnabla_\FF \ev = \UUa,
 	\quad
	\bnabla_{\UU_w} \ev = \UUa_w = -\Pa \nn - \nnu \bnabla \UUa \bcdot \nn
\ee
where this time the adjoint base flow $(\UUa,\Pa)^T$ is solution of the linear system 
\be 
	\bnabla \bcdot \UUa  =0,
 	\quad
  	-\bnabla \UUa \bcdot  \UU_b + \bnabla \UU_b^T \bcdot  \UUa
	- \bnabla \Pa - \nnu \bnabla^2 \UUa = \bnabla_\UU \ev,
\label{eq:adjBF-om}
\ee
with boundary conditions  
$\UUa=\00$ on $\Gamma_{in} \cup \Gamma_w \cup \Gamma_c$,
$\partial_y \Ua=0, \Va=0$ on $\Gamma_{lat}$, 
and $\Pa \nn+\nnu\bnabla \UUa \bcdot \nn + \UUa(\UU_b\bcdot\nn) + \uua(\bar\uu \bcdot \nn)=\00$ on $\Gamma_{out}$.
It is possible to relate the  sensitivity of a given eigenvalue and the individual sensitivities of its growth rate and pulsation, 
$\delta \sigma_{r,i} = (\bnabla_{\UU} \sigma_{r,i} \,|\, \bdelta\UU)$ 
according to
$\bnabla_{\UU} \grow =  \Real\{\bnabla_{\UU} \ev\}$ and
$\bnabla_{\UU} \puls = -\Imag\{\bnabla_{\UU} \ev\}$. 
The same relations hold for sensitivity to forcing.

\subsection{Numerical method} \label{sec:num}

All calculations are performed using the finite element software \textit{FreeFem++} to generate a two-dimensional triangulation in the computation domain $\Omega$ shown in figure~\ref{fig:geom_and_domain}, of dimensions
$-50 \leq x \leq 175$, $-30 \leq y \leq 30$, with the center of cylinder  located at $x=0, y=0$.
Bold lines indicate boundaries. 
The mesh density increases from the outer boundaries towards the cylinder wall, in successive regions indicated by dash-dotted lines in figure~\ref{fig:geom_and_domain}.
The resulting mesh has 246083 triangular elements.
Variational formulations associated to the equations to be solved are spatially discretized using 
P2 and P1 Taylor-Hood elements for velocity and pressure respectively.

Base flows are computed using an iterative Newton method to solve equations~(\ref{eq:NSsteady}), convergence being reached when the residual is smaller than $10^{-12}$ in $L^2$ norm.
The eigenvalue problem~(\ref{eq:evp}) is solved using an implicitly restarted Arnoldi method.
Adjoint base flows involved in the calculation of recirculation length sensitivity and eigenvalue sensitivity are obtained by inverting the simple linear systems (\ref{eq:adjBF-lc}) and (\ref{eq:adjBF-om}).

Convergence was checked by calculating steady-state base flows  with different meshes.
Reducing the number of elements by 21\%, the recirculation length varied by 0.2\% or less over the range of Reynolds numbers  $30 \leq \Rey \leq 120$.
Values of drag coefficient and recirculation length over this same $\Rey$ range
are given in table~\ref{tab:valid} together with results from the literature.
Compared to \citet{Gia07}, the maximum relative difference on $\rl$ was 2\%, of the same order as the values they report,
while $C_D$ differed by less than 1.7\% from values computed by  \citet{hen95}.
From linear stability calculations, the onset of instability characterized by $\grow=0$ was found to be $\Rey_c = 46.6$, in good agreement with the values reported in the literature.
Also, the frequency $\puls$ of the most unstable global mode differed by less than 1.2\% from results of \citet{Gia07}.

\begin{table} \small
  \begin{center}
\def~{\hphantom{0}}
  \begin{tabular}[]{l ccccc c ccccc c c}
                            & \multicolumn{5}{c}{$C_D$}             & & \multicolumn{5}{c}{$\rl$}              &   $\Rey_c$\\ 
  & $20$  & $40$  & $60$  & $100$ & $120$ & & $20$  & $40$  & $60$  & $100$ & $120$  \\ 
 \cline{2-6} \cline{8-12} \cline{14-14} 
\citet{hen95}             & 2.06 & 1.54 &  1.31  & 1.08 & 1.01    & &&&&&                                    &      \\
\citet{zie97}      &&&&&                                   & & 0.94  & 2.28  & 3.62  & 6.30  &        &      \\
\citet{Gia07}       & 2.05 & 1.54 &        &      &         & & 0.92  & 2.24  & (3.6) & (6.2) & (7.5)  & 46.7 \\
\citet{sip07}           &&&&&                                   & &&&&&                                    & 46.6 \\
\citet{mar08cyl}    &&&&&                                   & &&&&&                                    & 46.8 \\
Present study               & 2.04 & 1.52 &  1.30  & 1.07 & 1.00    & & 0.92  & 2.25  & 3.57  & 6.14  & 7.42   & 46.6 \\
  \end{tabular}
  \caption{\small 
  Drag coefficient $C_D$ and recirculation length $\rl$ for different values of $\Rey$, and critical Reynolds number $\Rey_c$. Bracketed numbers are estimated from a figure.}
  \label{tab:valid}
  \end{center}
\end{table}

\section{Results} 
\label{sec:results}

In this section we consider subcritical and supercritical Reynolds numbers, $40 \leq \Rey \leq 120$, and  focus on the steady-state recirculation length.
Its sensitivity to flow modification and to control is  presented in section~\ref{sec:sensitivity},
and examples of control configurations which  reduce $\rl$ are  detailed in section~\ref{sec:ctrl}.
Stability properties are discussed later in section~\ref{sec:stability}.

\subsection{Sensitivity of recirculation length} \label{sec:sensitivity}

\begin{figure}
  \vspace{0.2 cm}
  \centerline{
  \begin{overpic}[height=11 cm,tics=10]{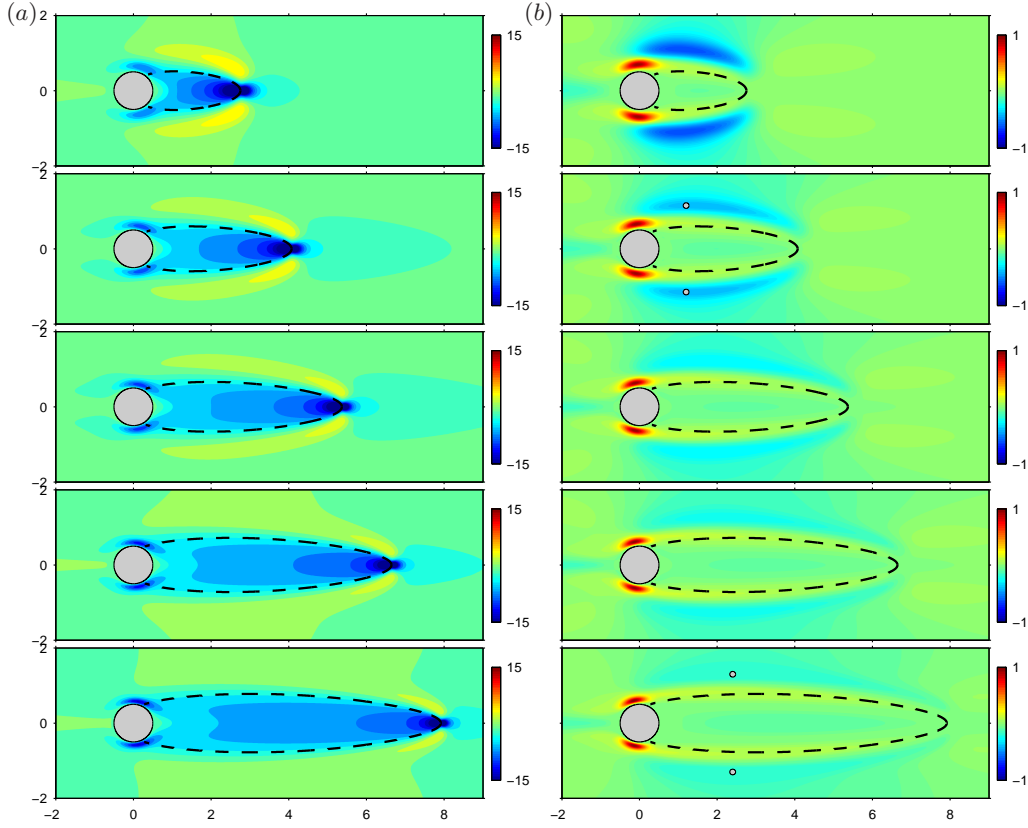}
     \put(-2.5,80) {$(a)$} 
     \put(49.,80) {$(b)$}
  \end{overpic}
  \vspace{0.2 cm}
  }
  \caption{
  (a) Normalized sensitivity of recirculation length to  bulk forcing in the streamwise direction, $\nabla_{F_x} \rl/\rl$. 
  (b)  Normalized effect of a small control cylinder of diameter $d=0.10D$ on recirculation length, $\delta \rl/\rl$.
From top to bottom: $\Rey=40, 60, 80, 100, 120$.
The dashed line is the steady-state base flow separatrix.
Black circles show the locations of control cylinders 
for configuration B discussed in section~\ref{sec:ctrl}.
 }
  \label{fig:sensit_l_F}
\end{figure}

Figure~\ref{fig:sensit_l_F}$(a)$ shows the sensitivity of recirculation length to bulk forcing in the streamwise direction,  
$\nabla_{F_x} \rl = \bnabla_{\FF} \rl \bcdot \ex$.
By construction, sensitivity analysis predicts that forcing has a large effect on $\rl$ in regions where sensitivity is large.
To be more specific, $\rl$ can be reduced by forcing along $\ex$   
in regions of negative sensitivity $\nabla_{F_x} \rl<0$:  in the recirculation region (in particular close to the reattachment point), and near the 
sides of the cylinder just upstream of the separation points; 
$\rl$ can  also be reduced by forcing along $-\ex$ in regions of positive sensitivity $\nabla_{F_x} \rl>0$: 
at the outer 
sides of the recirculation region (in particular close to the reattachment point).

\begin{figure}
  \def\thisfigytop{25} 
  \def\thisfigybot{-3}   
  \vspace{0.9cm}
  \centerline{
  \begin{overpic}[width=10.5cm, tics=10]{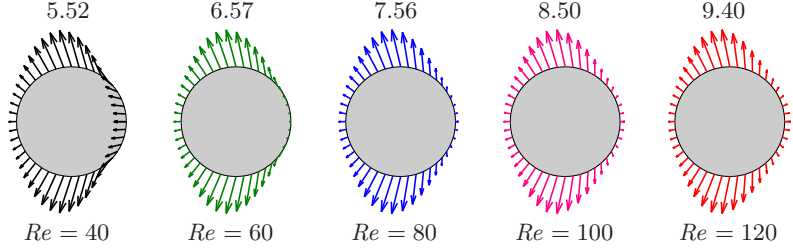}
    \put( 5,    \thisfigytop) {\footnotesize $5.52$}
	\put(25.75, \thisfigytop) {\footnotesize $6.57$}     
	\put(46.50, \thisfigytop) {\footnotesize $7.56$}     
	\put(67.25, \thisfigytop) {\footnotesize $8.50$}     
    \put(88,    \thisfigytop) {\footnotesize $9.40$}
    \put(2,     \thisfigybot) {\footnotesize $\Rey=40$}
	\put(22.75, \thisfigybot) {\footnotesize $\Rey=60$}     
	\put(43.50, \thisfigybot) {\footnotesize $\Rey=80$}     
	\put(64.25, \thisfigybot) {\footnotesize $\Rey=100$}     
    \put(85,    \thisfigybot) {\footnotesize $\Rey=120$}
  \end{overpic}
  }
  \vspace{0.4cm}
  \caption{  Sensitivity of recirculation length to wall actuation $\bnabla_{\UU_w} \rl$. 
  Flow is from left to right.
  Numbers  correspond to the $L^\infty$ norm of 
  $\bnabla_{\UU_w} \rl/\rl$.
  }
\label{fig:sensit_l_Uw-ntxy}
\end{figure}

\begin{figure}
  \vspace{0.3 cm}
  \centerline{
  \begin{overpic}[height=3.5 cm,tics=10]{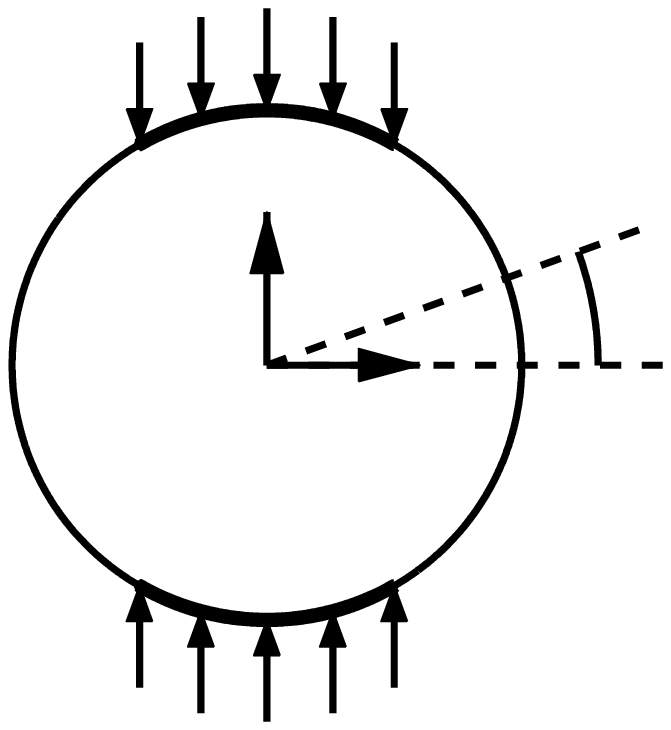}
 	\put(45, 37)  {\footnotesize $x$} 
	\put(23, 59)  {\footnotesize $y$} 
	\put(85, 54)  {\footnotesize $\theta$} 
    \put(-8, 90)  {\footnotesize $\delta V_w$} 
  \end{overpic} 
  }
  \caption{
  Sketch of control configuration W.
  }
\label{fig:sketch_wall_forcing}
\end{figure}

As mentioned in section~\ref{sec:smallctrlcyl}, it is convenient in an experiment 
to use a small, secondary  cylinder as a simple passive control device which produces a force which aligned with the local flow direction and depending non-linearly on the local velocity, as given by (\ref{eq:SmallCylForce}).
The effect $\delta \rl$ of such a control cylinder 
was computed from $\bnabla_{\FF} \rl$ according to (\ref{eq:SmallCylEffectRL}). 
Results for a control cylinder of diameter $d=0.10D$ are shown in figure~\ref{fig:sensit_l_F}$(b)$. 
The recirculation length increases ($\delta\rl>0$) when a control cylinder is located on the sides of the main cylinder upstream of the separation point , or in the shear layers; 
it decreases ($\delta\rl<0$) when a control cylinder is located farther away on the sides of the shear layers.
The minimal value of $\delta \rl/\rl$ becomes less negative as $\Rey$ increases, meaning that a control cylinder of constant diameter becomes gradually less effective at reducing $\rl$.
Interestingly,   regions associated with $\rl$ reduction  correspond qualitatively well to regions where vortex shedding was suppressed in the experiments of  \citet{Stry90} (figure 20 therein) for the same diameter ratio. This point is further discussed in section~\ref{sec:stability}.

Figure~\ref{fig:sensit_l_Uw-ntxy} shows the sensitivity of recirculation length to steady wall actuation, $\bnabla_{\UU_w} \rl$.
Since the variation of $\rl$ is given by the inner product $\delta \ev = \langle \bnabla_{\UU_w} \ev \,|\, \bdelta \UU_w \rangle$, wall control oriented along the arrows increases $\rl$. 
In particular, the recirculation length is increased by wall blowing where arrows point towards the fluid domain, and by wall suction where arrows point inside the cylinder.
The numbers above each plot correspond to the $L^\infty$ norm of the rescaled sensitivity field $\bnabla_{\UU_w} \rl/\rl$.
This norm increases (roughly linearly) with $\Rey$, indicating that the relative control authority  is increasing.
The shape of the sensitivity field does not vary much with $\Rey$, and reveals that
the most efficient way to reduce $\rl$ is to use wall suction at the top and bottom sides of the cylinder, in a direction close to wall normal.

\subsection{Control of the recirculation length} \label{sec:ctrl}

The sensitivity fields obtained in the previous section can be used to control recirculation length.
To illustrate this process, two control configurations predicted to reduce $\rl$ are tested at two representative Reynolds numbers $\Rey=60$ and 120, and for different control amplitudes: 
\begin{description}
\item - configuration B (``bulk''; sketched in figure~\ref{fig:sensit_l_F}): volume forcing with two control cylinders located symmetrically close to the point where  $\rl$ reduction is predicted to be maximal, at 
$\xx^* = (x^*,\pm y^*) = (1.2,\pm 1.15)$ at $\Rey=60$   and
$\xx^* = (2.4,\pm 1.3)$  at $\Rey=120$;
\item - configuration W  (``wall''; sketched in figure~\ref{fig:sketch_wall_forcing}): vertical wall  suction at the top and bottom sides, $\pi/3 \leq |\theta| \leq 2\pi/3$, with velocity $\delta V_w$.
\end{description}

\begin{figure}
  \psfrag{dU/dF}[t][][1][0]{$\delta F,\,|\delta V_w|$}
  \psfrag{F} [t][][1][0]{$\delta F$}
  \psfrag{lc}[br][l][1][-90]{$\rl\,\,\,$}
  \psfrag{llc}[b][][1][-90]{}
  \psfrag{d=0.05D} [][][1][0]{\footnotesize $d=0.05D$}
  \psfrag{d=0.1D} [][][1][0]{\footnotesize $d=0.10D$}
  \centerline{
  \begin{overpic}[height=5 cm,tics=10]{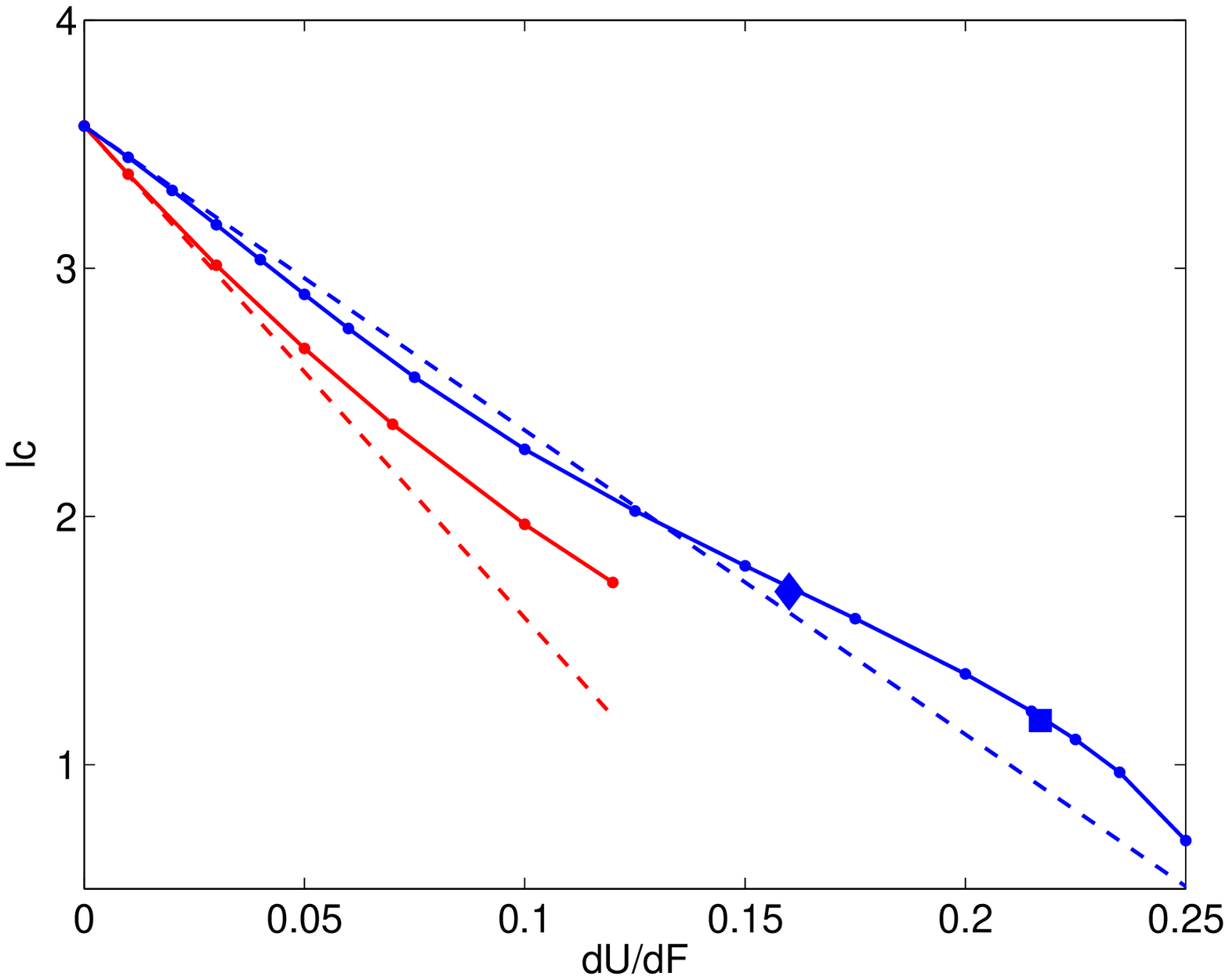}
   	\put(-5,75) {$(a)$} 
    \put(25,35) {\footnotesize \textcolor{red}{W}} 
 	\put(55,40) {\footnotesize \textcolor{blue}{B}} 
  \end{overpic} 
  \hspace{0.4 cm}
  \begin{overpic}[height=5 cm,tics=10]{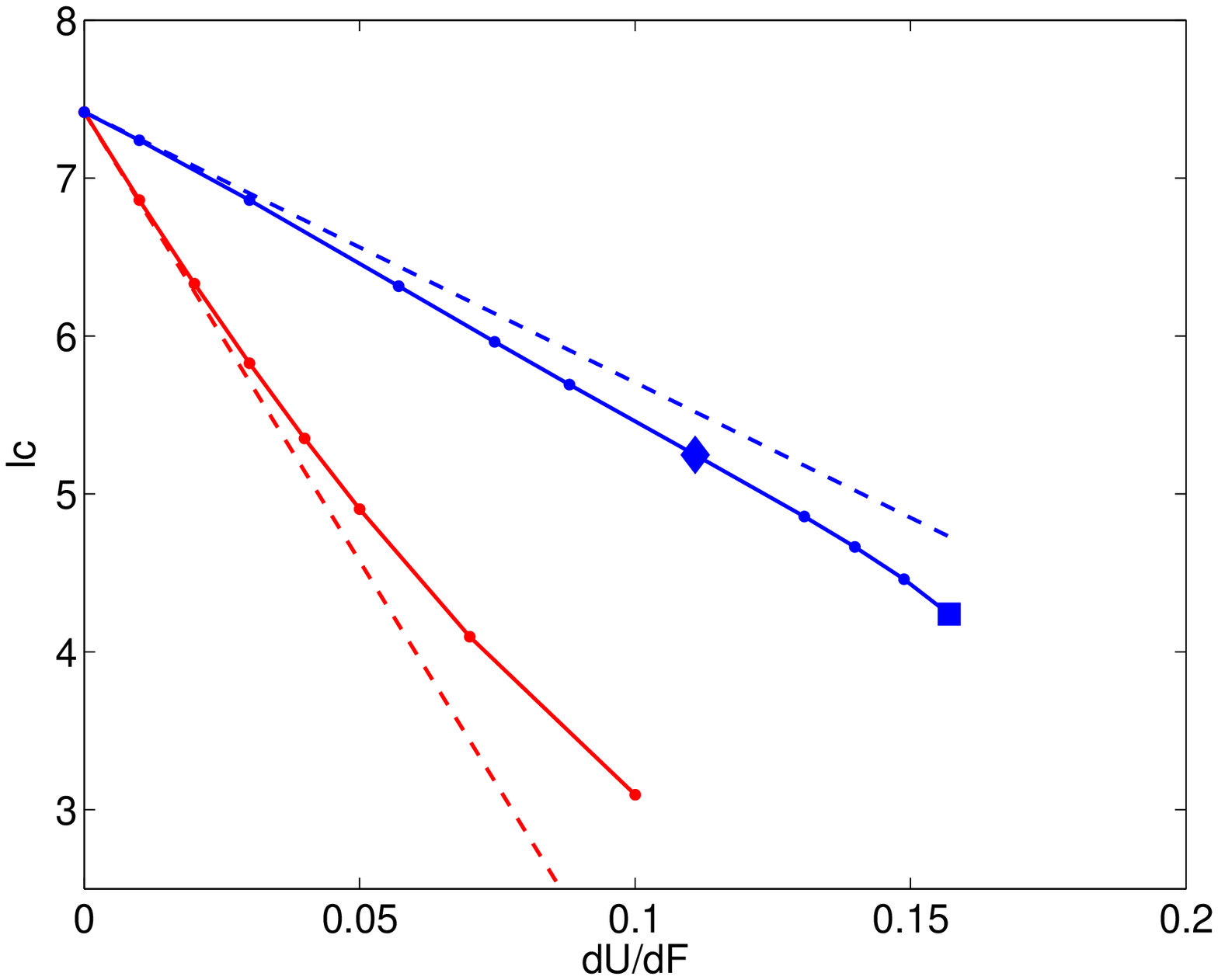}
     \put(-5,75) {$(b)$}
     \put(17,35) {\footnotesize \textcolor{red}{W}} 
 	 \put(65,45) {\footnotesize \textcolor{blue}{B}} 
  \end{overpic} 
  }
  \caption{
  Variation of recirculation length with control amplitude 
($\delta F$ in configuration B, $|\delta V_w|$ in configuration W).
    $(a)$ $\Rey=60$, $(b)$ $\Rey=120$.
    Dashed lines show predictions from sensitivity analysis,
solid lines are non-linear results.
Symbols correspond to control cylinders of diameter 
 $d=0.05 D$ ($\blacklozenge$)
and $d=0.10 D$ ($\scriptstyle\blacksquare$).
  }
\label{fig:sensit_dl_F}
\end{figure}

\begin{figure}
\vspace{0.7cm}
\hspace{0.2cm}
  \centerline{
  \begin{overpic}[width=13 cm,tics=10]{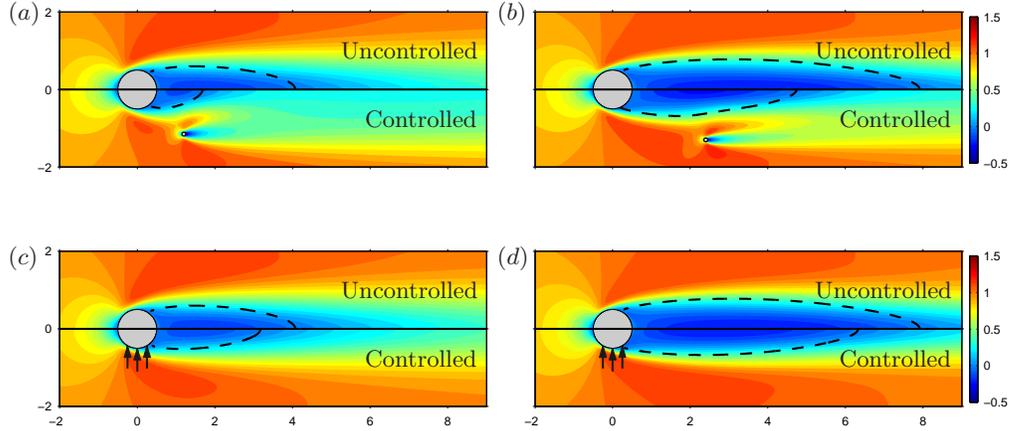}
     \put(-3,42) {$(a)$} 
     \put(47,42) {$(b)$} 
     \put(-3,17) {$(c)$} 
     \put(47,17) {$(d)$} 
     \put(31,38) {\small {Uncontrolled}}
     \put(33.5,31) {\small {Controlled}}
     \put(31,13.5) {\small {Uncontrolled}}
     \put(33.5,6.5)  {\small {Controlled}}
     \put(79.5,38) {\small {Uncontrolled}}
     \put(82,31) {\small {Controlled}}
     \put(79.5,13.5) {\small {Uncontrolled}}
     \put(82,6.5)  {\small {Controlled}}
  \end{overpic}
  }
  \caption{
Streamwise velocity of uncontrolled and controlled steady-state base flow.
  Left: $\Rey=60$; right: $\Rey=120$.
	$(a)$-$(b)$ configuration B: two control cylinders of diameter $d=0.10 D$ located at $(x^{*},\pm y^{*})$;
   $(c)$-$(d)$ configuration W:  suction of amplitude $\delta V_w=0.05$ and $0.03$ respectively.
   }
\label{fig:modifiedBF}
\end{figure}

Figure~\ref{fig:sensit_dl_F} 
shows  the recirculation length variation $\delta \rl$ for these  configurations. 
In addition to predictions from sensitivity analysis (dashed lines), this figure shows  non-linear results (solid lines) obtained by computing the actual controlled flow and its recirculation length for each configuration and each amplitude.
In configuration B, the effect of control cylinders is modelled by the volume force (\ref{eq:SmallCylForce}); 
in configuration W, wall actuation is implemented as a velocity boundary condition with uniform profile.
As expected, $\rl$ is reduced, and the agreement between sensitivity analysis and non-linear results is excellent, with slopes matching at zero-amplitude, whereas non-linear effects appear for larger amplitudes. 
Non-linear simulations indicate that at $\Rey=60$, controlling with two cylinders of diameter $d=0.10 D$ reduces $\rl$ by more than 65\%;
a reduction of about 50\% is achieved with control cylinders of diameter $0.05D$, and with wall suction of  intensity $\delta V_w = 0.12$.
At $\Rey=120$, control cylinders of diameter $d=0.10 D$ reduce $\rl$ by more than 40\%;
a reduction of about 30\% is achieved with control cylinders of diameter $0.05D$, and with wall suction of  intensity $\delta V_w = 0.04$.

Figure~\ref{fig:modifiedBF} shows examples of controlled flows, 
illustrating how the recirculation region is shortened.
In configuration B, 
control cylinders of diameter $d=0.10 D$  located at $(x^*,\pm y^*)$ make the flow deviate and accelerate. As a result, the streamwise velocity increases between the two control cylinders 
in a long region extending far downstream, and the reattachment point moves upstream.
In configuration W, 
fluid is sucked at the cylinder sides, which brings fluid with high streamwise velocity closer to the wall region, which in turn makes the recirculation region shorter.
%

\section{Effect on linear stability} \label{sec:stability}

In the previous section, sensitivity analysis was performed on the steady-state base flow. 
It provided information on the sensitivity of the recirculation length and allowed to design efficient control strategies.
These results are relevant only if the controlled flow is stable.
Indeed, when increasing the Reynolds number above its critical value $\Rey_c$, the  cylinder flow becomes linearly unstable, with a Hopf bifurcation leading to unsteady vortex shedding;
 one should therefore investigate whether the flow is stabilised by the control.
In the subcritical regime $\Rey < \Rey_c$, one should also check that the flow is not destabilised by the control.

\begin{figure}
  \vspace{0.3 cm}
  \psfrag{omi}[c][t][1][-90]{$\grow$}
  \psfrag{growth rate}[c][t][1][-90]{}
  \psfrag{F} [t][][1][0]{$\delta F$}
  \psfrag{dU/dF}[t][][1][0]{$\delta F, \, |\delta V_w|$}
  \centerline{   
  \begin{overpic}[height=4.7 cm,tics=10]{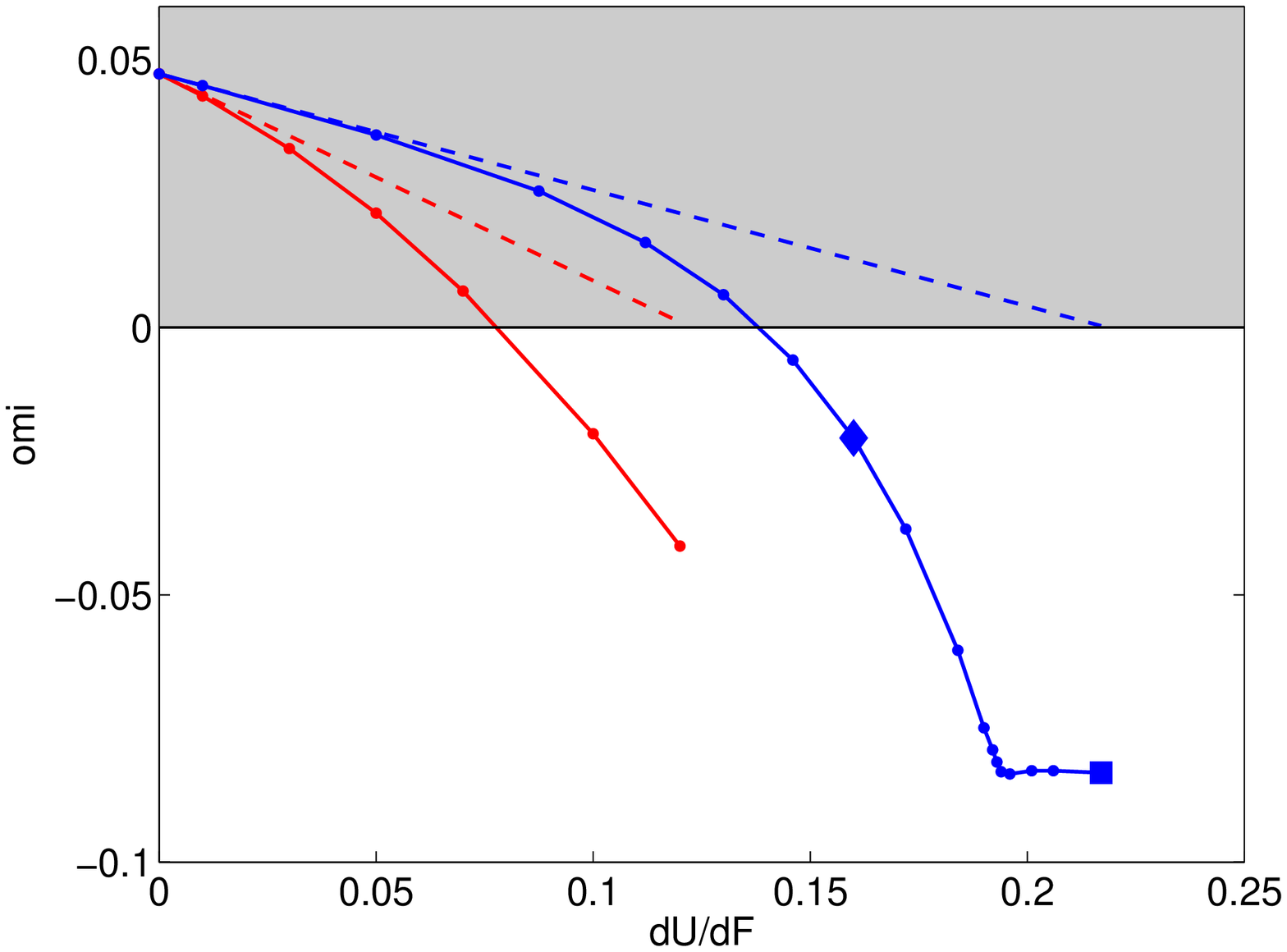}
  	\put(-2,70) {$(a)$} 
  	\put(25,67) {\footnotesize Mode 1} 
  	\put(42,33) {\footnotesize \textcolor{red}{W}} 
  	\put(73,33) {\footnotesize \textcolor{blue}{B}} 
  \end{overpic}
  \hspace{0.3 cm}
  \begin{overpic}[height=4.7 cm,tics=10]{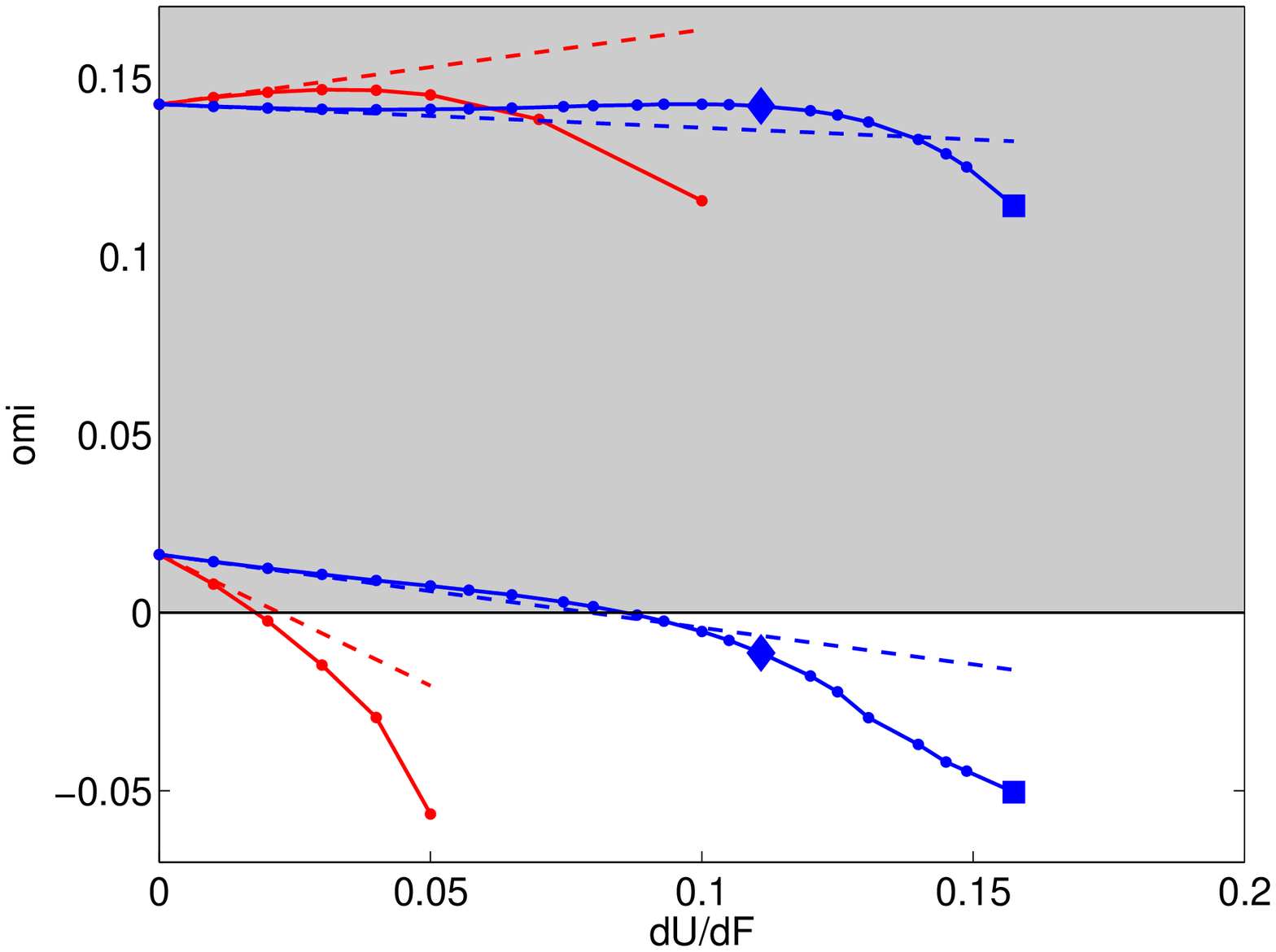}
  	\put(-2,71) {$(b)$}
  	\put(16,60) {\footnotesize Mode 1} 
  	\put(16,33) {\footnotesize Mode 2} 
  	\put(46,55) {\footnotesize \textcolor{red}{W}} 
  	\put(82,55) {\footnotesize \textcolor{blue}{B}} 
  	\put(22,15) {\footnotesize \textcolor{red}{W}} 
  	\put(82,15) {\footnotesize \textcolor{blue}{B}} 
  \end{overpic}
  }
  \caption{
   Variation of leading growth rates with control amplitude.
   $(a)$ $\Rey=60$, $(b)$~$\Rey=120$. 
Same notations as in figure~\ref{fig:sensit_dl_F}. 
  }
\label{fig:sensit_domi_F}
\end{figure}

At $\Rey=60$, one pair of complex conjugate eigenvalues (``mode 1'') associated with the von K\'arm\'an street  is unstable.
Figure~\ref{fig:sensit_domi_F}$(a)$ 
shows the variation of the leading growth rate  
$\sigma_{1,r}$
with control amplitude.
Both configurations B and W have a stabilising effect on the leading global mode. 
Full restabilisation is achieved with two control
cylinders of diameter $d \simeq 0.04 D$, 
or with wall suction of intensity $\delta V_w \simeq 0.08$.
These control configurations do not destabilise other eigenmodes for any of the amplitudes tested.

\begin{figure}
\def\thisfigdx{62}
\vspace{0.2 cm}
\hspace{1.7 cm}
  \centerline{
  \begin{overpic}[width=12 cm,tics=10]{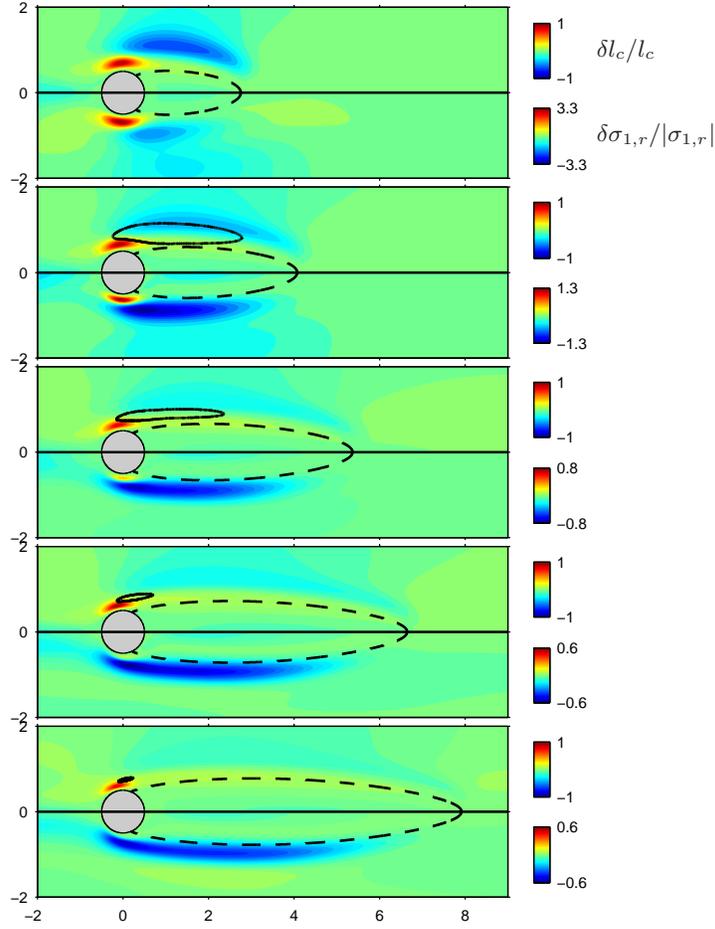}  
     \put(\thisfigdx, 92){\small $\delta\rl/\rl$}
     \put(\thisfigdx, 83){\small $\delta\sigma_{1,r} / |\sigma_{1,r}|$}
  \end{overpic}
  }
  \caption{
  Normalized effect of a small control cylinder ($d=0.10D$)
  on the recirculation length, $\delta\rl/\rl$           (upper half of each panel), and on the most unstable growth rate, $\delta \sigma_{1,r} / |\sigma_{1,r}|$ (lower half).
From top to bottom: $\Rey=40, 60, 80, 100, 120$.
  The stabilising contour  $\delta \sigma_{1,r} / \sigma_{1,r}=-1/2$ for two symmetric control cylinders is reported on the upper half plots for $\Rey > \Rey_c$.
  }
\label{fig:sensit_omi_F}
\end{figure}

A second pair of eigenvalues becomes unstable (``mode 2'') at $\Rey \simeq 110$ \citep{ver11}.
Figure~\ref{fig:sensit_domi_F}$(b)$ shows the variation of $\sigma_{1,r}$ and $\sigma_{2,r}$
with control amplitude at $\Rey=120$.
Bulk forcing has a  stabilising effect on both unstable modes, but does not achieve full restabilisation. 
The straightforward strategy which allowed to stabilise the flow at $\Rey=60$,
namely placing control cylinders where the sensitivity analysis predicts they have the largest reducing effect on $\rl$, is therefore not  successful at $\Rey=120$.
Wall forcing has a  stabilising effect on mode 2 
but a destabilising effect on mode 1 for reasonable control amplitude, and again the flow is not restabilised.

To investigate why control configurations which shorten the recirculation region also have a stabilising effect at $\Rey=60$, but not at $\Rey=120$, it is interesting to consider the sensitivity of the leading eigenvalue.
Figure~\ref{fig:sensit_omi_F} 
compares the effect of a small control cylinder 
($d=0.10D$)
on $\rl$, calculated from (\ref{eq:SmallCylEffectRL}) and already shown in figure~\ref{fig:sensit_l_F}$(b)$, 
and its effect on $\sigma_{1,r}$,
calculated from  (\ref{eq:SmallCylEffectEV}).
At $\Rey=40$ and 60, 
$\delta\rl$ and $\delta\sigma_{1,r}$ have very similar spatial  structures. 
This means that a small control cylinder, when  located where it reduces the recirculation length, almost always has a stabilising effect on the most unstable global mode.
However, this similarity gradually disappears as $\Rey$ increases.
Regions where $\delta\rl$ and $\delta\sigma_{1,r}$ have opposite signs 
grow in size, and at $\Rey=120$ they extend along the whole shear layers, both inside and outside the recirculation region.
To ease comparison, the contour where two symmetric control cylinders ($d=0.10D$) render mode 1 just neutrally stable
(i.e. where $\delta \sigma_{1,r} = - \sigma_{1,r}/2$) is reported on the map of $\delta\rl$. 
At $\Rey=60$ this stabilising region overlaps with the region of recirculation length reduction, but   
as $\Rey$ increases it moves upstream towards the region of recirculation length increase.
In other words,  control cylinders located where they reduce $\rl$ are efficient in stabilising mode 1 at low $\Rey$,
 but gradually lose this ability at higher $\Rey$. 
In the latter regime, one may wonder if increasing the recirculation length is not a better way to stabilise the flow.
It must be pointed out that the stabilising region shrinks as $\Rey$ increases anyway, consistent with observations from \citet{Stry90} and \citet{mar08cyl}.

\begin{figure}
  \psfrag{x} [ ][ ][1][  0]{$x$}
  \psfrag{y} [r][l][1][-90]{$y$}
  \psfrag{dlc/lc}[][][1][-90]{$\displaystyle \frac{\delta\rl}{\rl} \quad$}
  \psfrag{dsigr/sigr}[][t][1][-90]{$\displaystyle \frac{\delta\sigma_{1,r}}{\sigma_{1,r}} \quad$}
  \psfrag{re}   [t][][1][0]{$\Rey$}
  \centerline{
  	\begin{overpic}[width=10cm,tics=10]{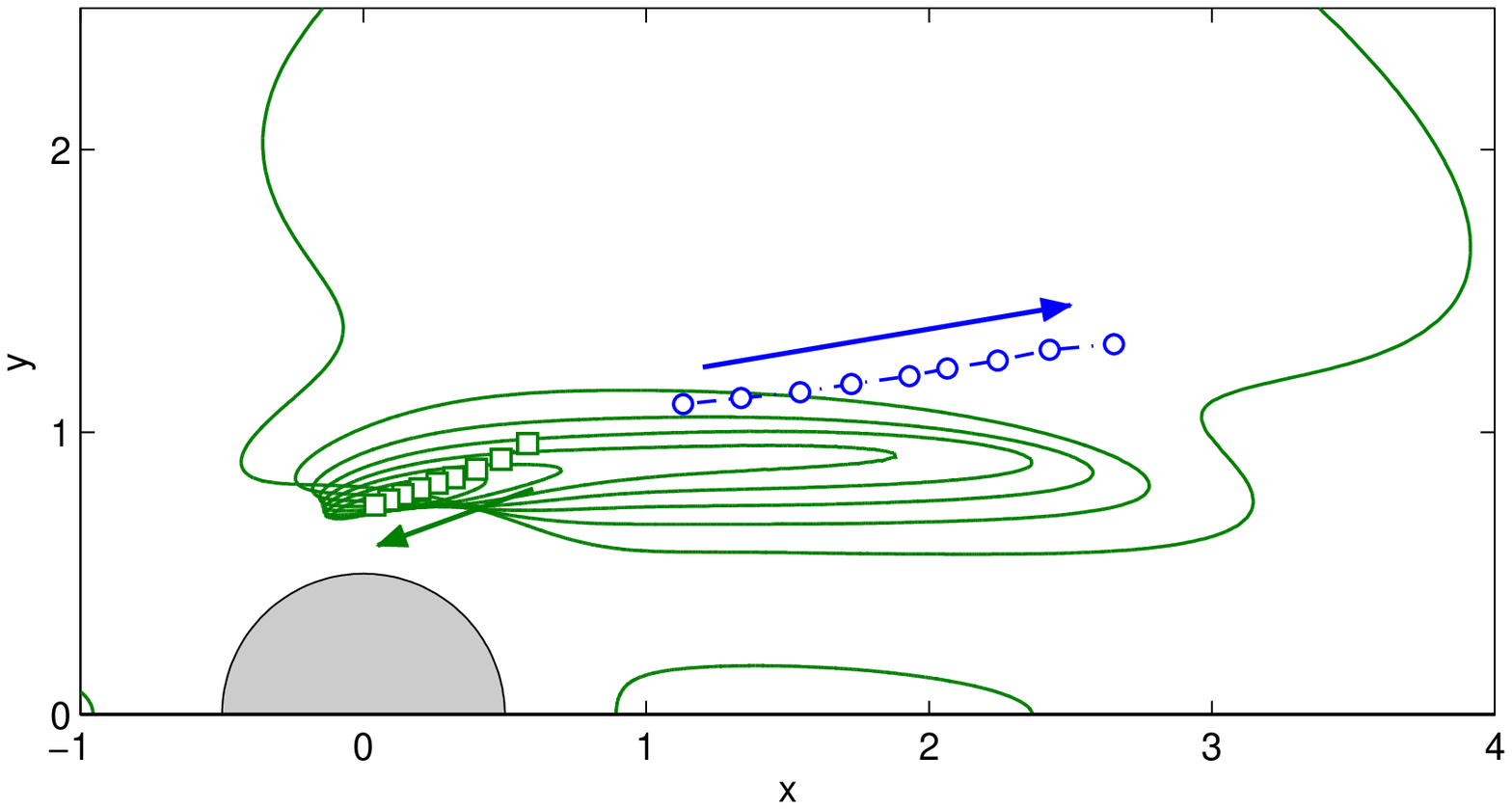}
		\put(-3,50) {$(a)$}
		\put(30,16) {\textcolor{mygreen}{\footnotesize $\xx_\sigma$}} 
  		\put(57,34) {\textcolor{blue}   {\footnotesize $\xx_l$}}
  	\end{overpic}
  }
  \vspace{0.4 cm}
  \centerline{ 
  	\begin{overpic}[height=5cm, tics=10]{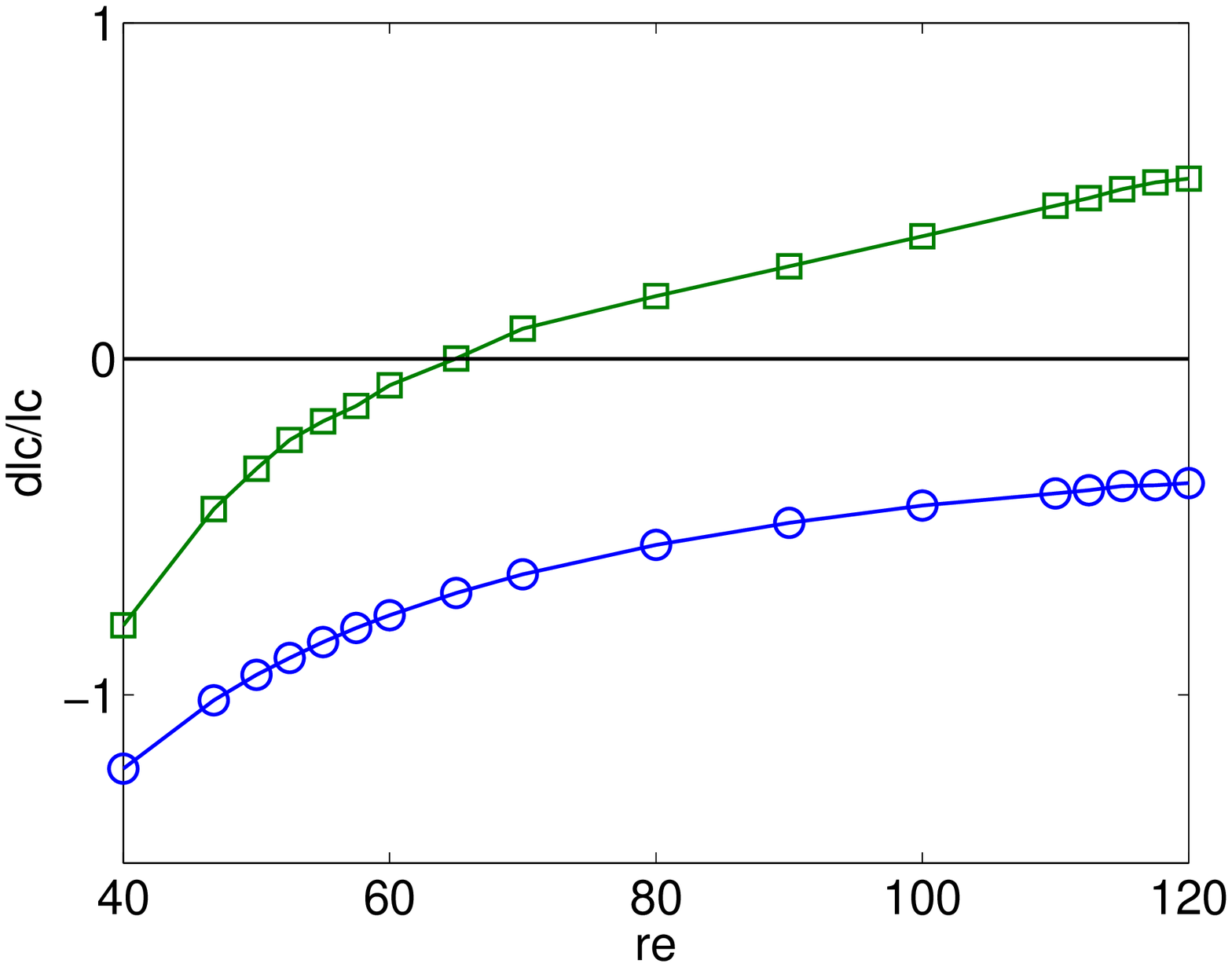}
    	\put(-4,73) {$(b)$}
    	\put(30,57) {\textcolor{mygreen}{\footnotesize $\xx_c = \xx_\sigma$}}
    	\put(57,30) {\textcolor{blue}   {\footnotesize $\xx_c = \xx_l$}}
  	\end{overpic}
  	\hspace{0.7 cm}
  	\begin{overpic}[height=5cm, tics=10]{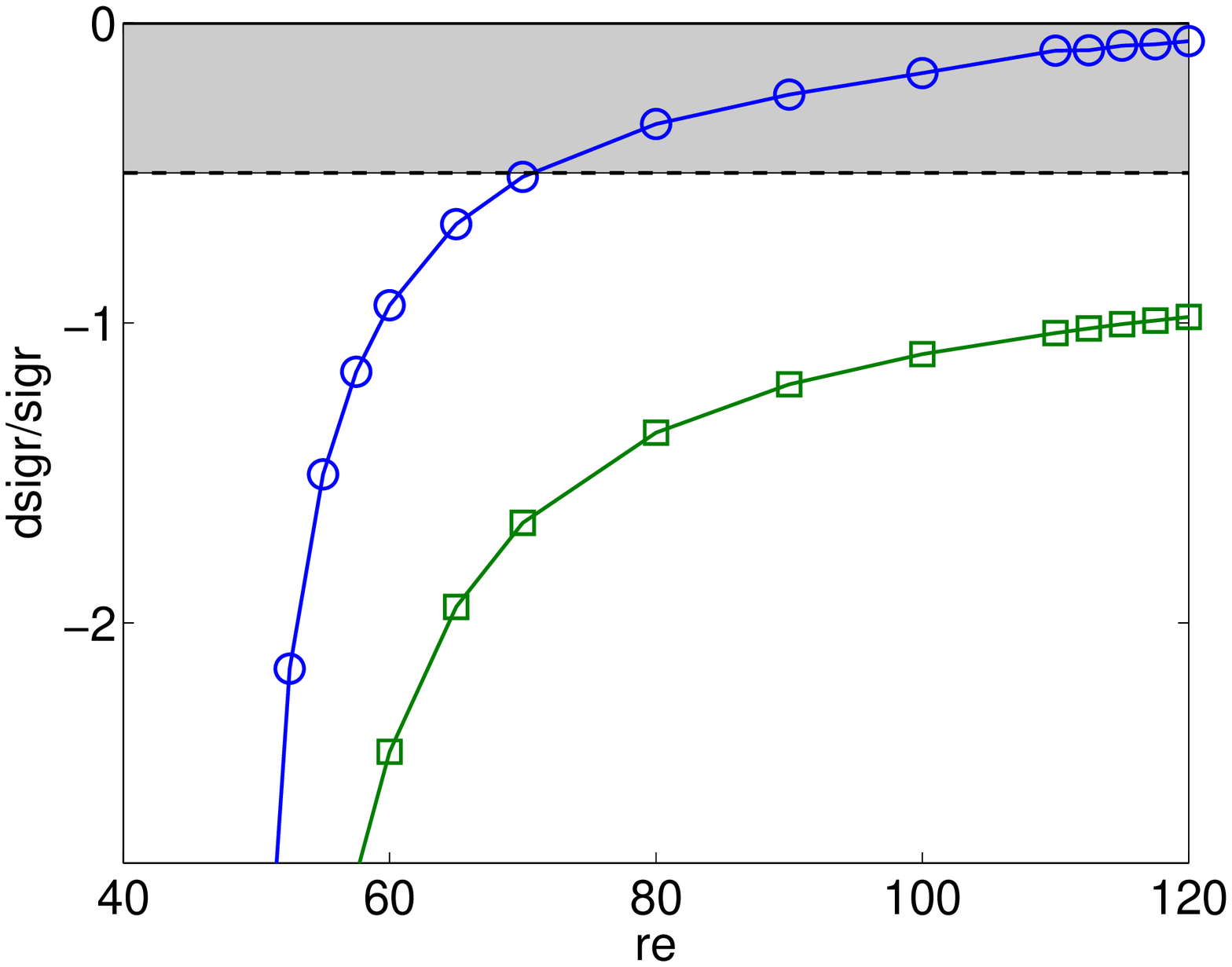}
    	\put(-5,73) {$(c)$} 
    	\put(89.5,66) {\textcolor{black}   {U}} 
    	\put(90,59) {\textcolor{black}   {S}}
    	\put(15,57) {\textcolor{blue}   {\footnotesize $\xx_c = \xx_l$}}
    	\put(45,32) {\textcolor{mygreen}{\footnotesize $\xx_c = \xx_\sigma$}}
  	\end{overpic}
  }
  \caption{ 
   $(a$) Locus of $\xx_l$          (where $-\delta\rl/\rl$ is maximal)
  and      $\xx_\sigma$ (where $-\delta\sigma_{1,r}/|\sigma_{1,r}|$ is maximal)
  for a small control cylinder ($d=0.10D$)
  at $\Rey=40, 50 \ldots 120$.
Solid lines show stabilising regions  $\delta \sigma_{1,r} / \sigma_{1,r} \leq -1/2$ for two symmetric control cylinders at $\Rey=50, 60 \ldots 120$.
 $(b)$-$(c)$ Effect of two symmetric control cylinders ($d=0.10D$) located at $\xx_l$ 
  or at $\xx_\sigma$: 
  sensitivity analysis prediction for the 
  normalized variation of $(a)$~recirculation length,
                          $(b)$~growth rate of mode 1.
  The unshaded area  
  corresponds to restabilisation of  mode 1. 
  }
  \label{fig:lose-correlation}
\end{figure}

All stabilising contours for $\Rey \geq 50$ are gathered  in figure~\ref{fig:lose-correlation}$(a)$. 
The characteristic shrinking of the stabilising region is confirmed.
Also shown for $\Rey \geq 40$ are the locus of two particular points:
$\xx_l$, where a control cylinder yields the largest $\rl$ reduction,
and 
$\xx_\sigma$, where a control cylinder has the maximal stabilising effect on mode 1. 
 It can clearly be observed that $\xx_l$ and $\xx_\sigma$ move in opposite directions, with $\xx_l$ eventually going outside the stabilising region. 
Quantitative values of $\delta \rl$ and $\delta \sigma_{1,r}$ are given  in figures~\ref{fig:lose-correlation}$(b)$-$(c)$.
With control cylinders located at $\xx_l$, a recirculation length reduction of more than 35\% can be achieved for any $\Rey \leq 120$. 
(With $\xx_c = \xx_\sigma$, the reduction is of course not as large, and the recirculation length actually increases when $\Rey \gtrsim 65$.)
As stressed previously, however, the corresponding controlled steady-state base flow will be observed only if stable, i.e. if $\delta \sigma_{1,r}/\sigma_{1,r} \leq 1/2$ (region labelled S). 
This is the case if $\xx_c = \xx_\sigma$ (at least for $\Rey \leq 120$).
But when control cylinders are located at $\xx_l$, although they do have a stabilising effect ($\delta \sigma_{1,r} \leq 0$), the latter is too small to restabilise the flow when $\Rey \gtrsim 70$.

\begin{figure}
  \def\thisfigytop{25} 
  \def\thisfigybot{-3}  
  \psfrag{Scal Prod}[][][1][0]{$\bnabla_{\UU_w} \rl \bcdot \bnabla_{\UU_w} \sigma_{1,r}$}  
  \psfrag{theta}[t][][1][0]{$\theta$}
  \psfrag{-pi}  [][][1][0]{$-\pi$}
  \psfrag{-pi/2}[][][1][0]{$-\pi/2$}
  \psfrag{00}   [][][1][0]{$0$}
  \psfrag{pi}   [][][1][0]{$\pi$}
  \psfrag{pi/2} [][][1][0]{$\pi/2$}
  \vspace{0.9 cm}
  \centerline{   
  \hspace{1 cm}   
  \begin{overpic}[width=10.5cm,tics=10]{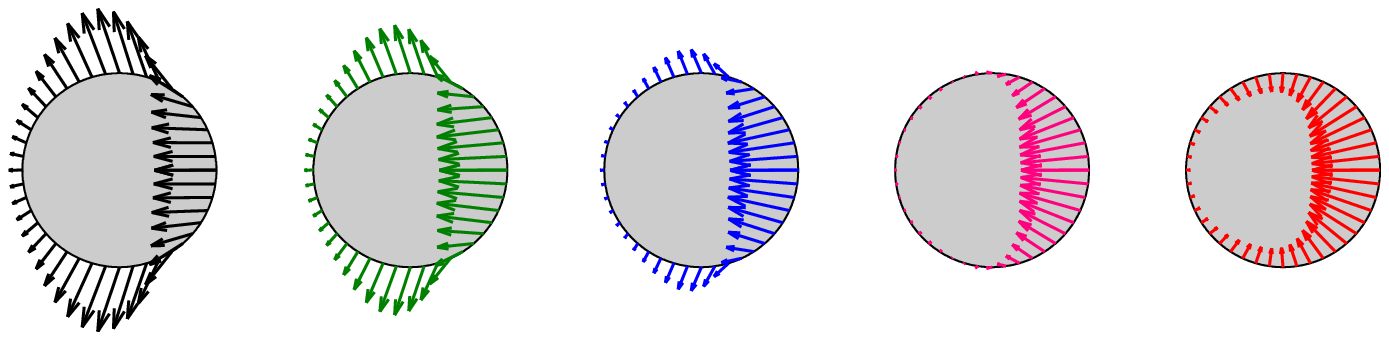}
    \put(-5,   25) {\footnotesize $(a)$}
    \put( 5,    \thisfigytop) {\footnotesize $0.281$}
	\put(25.75, \thisfigytop) {\footnotesize $0.186$}     
	\put(46.50, \thisfigytop) {\footnotesize $0.147$}     
	\put(67.25, \thisfigytop) {\footnotesize $0.117$}     
    \put(88,    \thisfigytop) {\footnotesize $0.091$}  
    \put(2,     \thisfigybot) {\footnotesize $\Rey=40$}
	\put(22.75, \thisfigybot) {\footnotesize $\Rey=60$}     
	\put(43.50, \thisfigybot) {\footnotesize $\Rey=80$}     
	\put(64.25, \thisfigybot) {\footnotesize $\Rey=100$}     
    \put(85,    \thisfigybot) {\footnotesize $\Rey=120$}  
  \end{overpic}
  }
  \vspace{0.9 cm}
  \centerline{ 
  \begin{overpic}[width=10cm,tics=10]{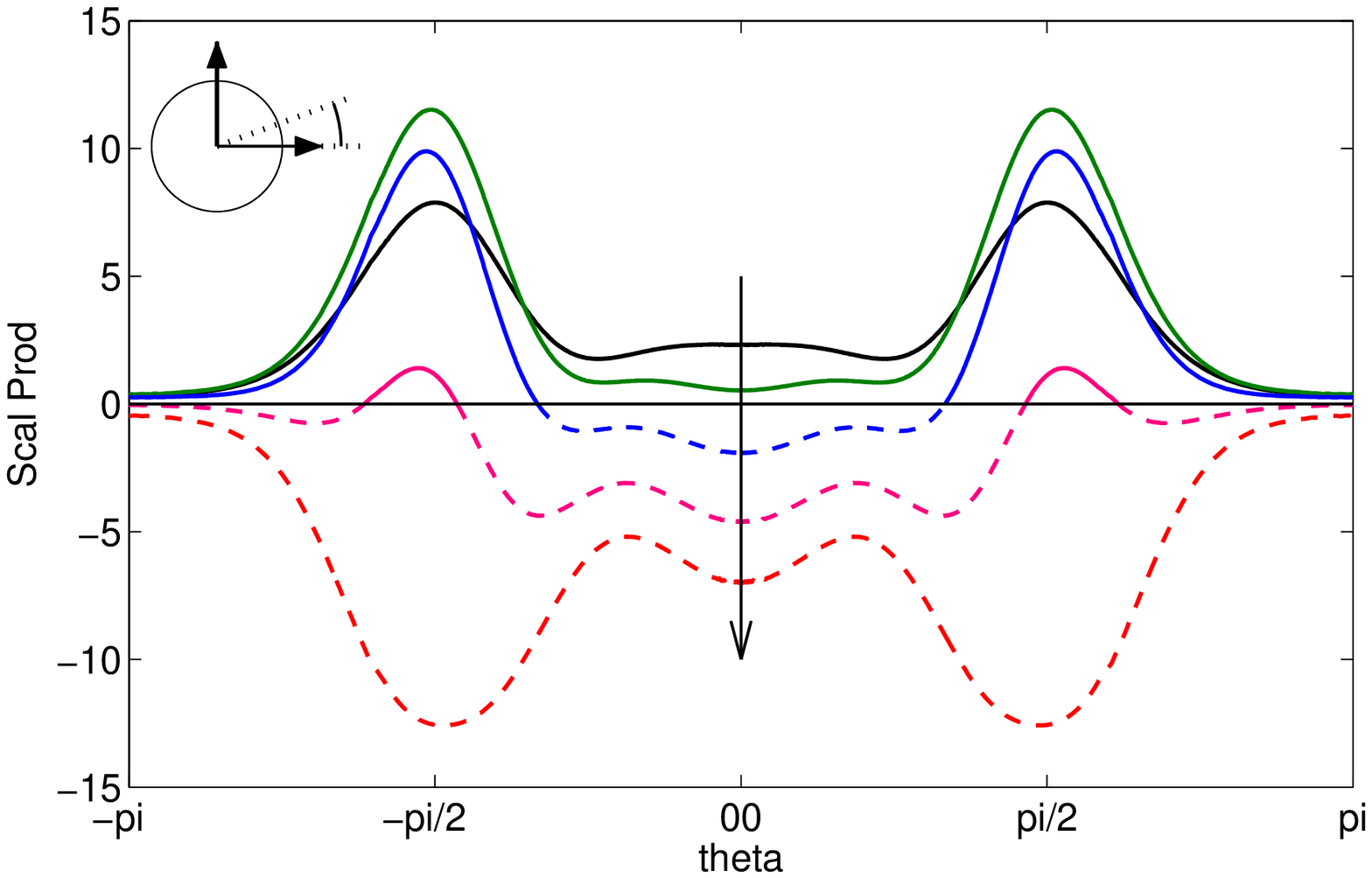}
     \put(-2,61) {\footnotesize $(b)$}
     \put(48,45.5) {\footnotesize $\Rey=40$}
     \put(48,13) {\footnotesize $\Rey=120$}
     \put(26,54) {\footnotesize $\theta$}
     \put(22,50) {\footnotesize $x$}
     \put(13,60) {\footnotesize $y$}
  \end{overpic}
  }
  \caption{ 
    $(a)$~Sensitivity of the leading growth rate to wall actuation
  $\bnabla_{\UU_w} \sigma_{1,r}$. 
  Flow is from left to right.
  Numbers  correspond to the $L^\infty$ norm of 
   $\bnabla_{\UU_w} \sigma_{1,r}/\sigma_{1,r}$.
  $(b)$~Pointwise inner product of $\bnabla_{\UU_w} \rl$  (figure \ref{fig:sensit_l_Uw-ntxy})
  and                        $\bnabla_{\UU_w} \sigma_{1,r}$ 
  along the cylinder wall.
  }
  \label{fig:D_Uw_evgrowth-lc-ScalProd}
\end{figure}

The same phenomenon is observed with sensitivity  to wall forcing.
Figure~\ref{fig:D_Uw_evgrowth-lc-ScalProd}$(a)$ shows the sensitivity of the most unstable growth rate to wall forcing.
It compares very well with the results  at $\Rey=60$ of 
\citet[figure 3 therein]{mar10-adjoint}.
Unlike the sensitivity of recirculation length (figure \ref{fig:sensit_l_Uw-ntxy}) which keeps more or less the same structure at all Reynolds numbers, $\bnabla_{\UU_w} \sigma_{1,r}$ varies substantially. 
In particular, at the top and bottom sides of the cylinder, wall-normal suction changes from largely stabilising to slightly destabilising.
This translates into $\bnabla_{\UU_w} \rl$ and $\bnabla_{\UU_w} \sigma_{1,r}$ being very similar at $\Rey=40$ but very different at $\Rey=120$. 
Accordingly, the pointwise inner product of these two fields shown in figure~\ref{fig:D_Uw_evgrowth-lc-ScalProd}$(b)$ is positive everywhere on the cylinder wall at $\Rey=40$ but negative everywhere at $\Rey=120$.
As a result, control configurations which shorten the recirculation length necessarily have a stabilising effect on mode 1 at low Reynolds numbers and a destabilising effect at higher Reynolds numbers.
It should be noted that at $\Rey=100$, it is still possible to reduce $\rl$ and at the same time have a stabilising effect on mode 1 by using wall suction in a narrow region near $\theta=\pm\pi/2$.

\section{Conclusions}
\label{sec:conclusion}

In this study, the sensitivity of recirculation length to steady forcing was derived analytically using a variational technique.
Linear sensitivity analysis was applied to the two-dimensional steady flow past a circular cylinder for 
both subcritical and supercritical Reynolds numbers  $40 \leq \Rey \leq 120$.
Regions of largest sensitivity were identified:
$\rl$ increases the most when small control cylinders are located close to the top and bottom sides of the main cylinder, and decreases the most when control cylinders are located farther downstream,  outside the shear layers;
regarding wall forcing, $\rl$ is most sensitive to normal blowing/suction at the sides of the cylinder. 
Validation against full non-linear Navier-Stokes calculations showed excellent agreement for small-amplitude control.

Using linear stability analysis, it was  observed that control configurations which 
reduce $\rl$ also have a stabilising effect on the most unstable eigenmode close to $\Rey_c$, both for bulk forcing and wall forcing. 
This property gradually disappears as $\Rey$ is increased.
This is explained by the spatial structures of the  sensitivities of $\rl$ and $\sigma_{1,r}$, which are very similar at low $\Rey$, but increasingly decorrelated at larger $\Rey$.
Therefore, reducing the base flow recirculation length is an appropriate control strategy to restabilise the flow at moderate Reynolds numbers. 
At larger $\Rey$, aiming for an increase of $\rl$ is actually more efficient.
In any case, one should keep in mind that results concerning  $\rl$ reduction and obtained from a sensitivity analysis performed on the steady-state base flow are relevant only when the  controlled flow is stable.

To better control the flow in the supercritical regime, the sensitivity of the mean flow recirculation length should be considered.
Not only would a method allowing to control the mean $\rl$ be interesting in itself, but targeting this important parameter of the mean state \citep{zie97,thi07}  could also help stabilise the flow.
 This work is in progress, but the difficulty of such an approach lies in that the mean flow and the fluctuations are non-linearly coupled, 
which prevents the derivation of a simple expression for the 
 sensitivity of the mean recirculation length. 
It would help, though, to determine 
 whether the mean flow recirculation length in separated flows should be reduced or increased in order to mitigate the instability.
This extension of the sensitivity analysis to the mean recirculation length could also include the effect of periodic excitation, a control strategy often used in turbulent flows \citep{Gre00,Gle02} and which can be interpreted as a mean-flow correction \citep{Sip12}.

This study confirms the high versatility of Lagrangian-based variational techniques, which allow to compute the sensitivity  of a great variety of quantities of interest in fluid flows.
The recirculation length appears as a simple and relevant macroscopic parameter in separated flows, and can be targeted to design original control strategies.
In the case of the cylinder flow, the fact that $\rl$ is a good proxy for flow stabilisation only up to a certain Reynolds number 
might be specific
 to the absolute nature of the instability in bluff body wakes \citep{mon88bluff}. These flows  are typical examples of ``oscillators'', and are appropriately described by global linear stability analysis \citep{Cho05}.
On the other hand, convectively unstable flows (or ``noise amplifiers''), such as separated boundary layers, usually  exhibit  large optimal transient growth (maximal energy amplification of an initial perturbation) and  large optimal gain (maximal  energy amplification from  harmonic forcing to asymptotic response for a globally stable flow)  as a result of  the non-normality of the linearized Navier-Stokes operator \citep{Cho05}.
One may wonder whether the recirculation length is  more directly and strongly related to instability in such convectively unstable flows,
and whether sensitivity analysis applied to $\rl$ would be efficient over a broader range of Reynolds numbers.
The next step of this work is  the application of a similar control strategy in wall-bounded separated boundary layer flows.

\bigskip
This work is supported by the Swiss National Science Foundation (grant no. 200021-130315).
The authors are also thankful to the French National Research Agency (project no. ANR-09-SYSC-001).

\bibliographystyle{jfm}
\bibliography{biblio_lc}

\begin{thebibliography}{28}
\expandafter\ifx\csname natexlab\endcsname\relax\def\natexlab#1{#1}\fi

\bibitem[Acrivos {\em et~al.\/}(1968)Acrivos, Leal, Snowden \& Pan]{Acr68}
{\sc Acrivos, A., Leal, L.~G., Snowden, D.~D. \& Pan, F.} 1968 Further
  experiments on steady separated flows past bluff objects. {\em Journal of
  Fluid Mechanics\/} {\bf 34}, 25--48.

\bibitem[Barkley {\em et~al.\/}(2002)Barkley, Gomes \& Henderson]{bar02}
{\sc Barkley, D., Gomes, M. Gabriela~M. \& Henderson, R.~D.} 2002
  Three-dimensional instability in flow over a backward-facing step. {\em
  Journal of Fluid Mechanics\/} {\bf 473}, 167--190.

\bibitem[Bewley {\em et~al.\/}(2001)Bewley, Moin \& Temam]{Bew01}
{\sc Bewley, T.~R., Moin, P. \& Temam, R.} 2001 {D}{N}{S}-based predictive
  control of turbulence: an optimal benchmark for feedback algorithms. {\em
  Journal of Fluid Mechanics\/} {\bf 447}, 179--225.

\bibitem[Bottaro {\em et~al.\/}(2003)Bottaro, Corbett \& Luchini]{Bot03}
{\sc Bottaro, A., Corbett, P. \& Luchini, P.} 2003 The effect of base flow
  variation on flow stability. {\em Journal of Fluid Mechanics\/} {\bf 476},
  293--302.

\bibitem[Brandt {\em et~al.\/}(2011)Brandt, Sipp, Pralits \& Marquet]{Bra11}
{\sc Brandt, L., Sipp, D., Pralits, J.~O. \& Marquet, O.} 2011 Effect of
  base-flow variation in noise amplifiers: the flat-plate boundary layer. {\em
  Journal of Fluid Mechanics\/} {\bf 687}, 503--528.

\bibitem[Chomaz({2005})]{Cho05}
{\sc Chomaz, J.-M.} {2005} {Global instabilities in spatially developing flows:
  Non-normality and nonlinearity}. {\em {Annual Review of Fluid Mechanics}\/}
  {\bf {37}}, {357--392}.

\bibitem[Corbett \& Bottaro(2001)]{cor01tcfd}
{\sc Corbett, P. \& Bottaro, A.} 2001 Optimal control of nonmodal disturbances
  in boundary layers. {\em Theoretical and Computational Fluid Dynamics\/} {\bf
  15}~(2), 65--81.

\bibitem[Finn(1953)]{fin53}
{\sc Finn, R.~K.} 1953 Determination of the drag on a cylinder at low reynolds
  numbers. {\em Journal of Applied Physics\/} {\bf 24}~(6), 771--773.

\bibitem[Giannetti \& Luchini(2007)]{Gia07}
{\sc Giannetti, F. \& Luchini, P.} 2007 Structural sensitivity of the first
  instability of the cylinder wake. {\em Journal of Fluid Mechanics\/} {\bf
  581}, 167--197.

\bibitem[Glezer \& Amitay(2002)]{Gle02}
{\sc Glezer, A. \& Amitay, M.} 2002 Synthetic jets. {\em Annual Review of Fluid
  Mechanics\/} {\bf 34}~(1), 503--529.

\bibitem[Greenblatt \& Wygnanski(2000)]{Gre00}
{\sc Greenblatt, D. \& Wygnanski, I.~J.} 2000 The control of flow separation by
  periodic excitation. {\em Progress in Aerospace Sciences\/} {\bf 36}~(7),
  487--545.

\bibitem[Henderson(1995)]{hen95}
{\sc Henderson, R.~D.} 1995 Details of the drag curve near the onset of vortex
  shedding. {\em Physics of Fluids\/} {\bf 7}~(9), 2102--2104.

\bibitem[Hill(1992)]{Hill92AIAA}
{\sc Hill, D.~C.} 1992 A theoretical approach for analyzing the restabilization
  of wakes. {\em AIAA 92-0067\/} .

\bibitem[Marquet \& Sipp(2010)]{mar10-adjoint}
{\sc Marquet, O. \& Sipp, D.} 2010 Active steady control of vortex shedding: an
  adjoint-based sensitivity approach. In {\em Seventh IUTAM Symposium on
  Laminar-Turbulent Transition\/} (ed. Philipp Schlatter \& Dan~S. Henningson),
  {\em IUTAM Bookseries\/}, vol.~18, pp. 259--264. Springer Netherlands.

\bibitem[Marquet {\em et~al.\/}(2008)Marquet, Sipp \& Jacquin]{mar08cyl}
{\sc Marquet, O., Sipp, D. \& Jacquin, L.} 2008 Sensitivity analysis and
  passive control of cylinder flow. {\em Journal of Fluid Mechanics\/} {\bf
  615}, 221--252.

\bibitem[Marquillie \& Ehrenstein(2003)]{Mar03}
{\sc Marquillie, M. \& Ehrenstein, U.} 2003 On the onset of nonlinear
  oscillations in a separating boundary-layer flow. {\em Journal of Fluid
  Mechanics\/} {\bf 490}, 169--188.

\bibitem[Meliga {\em et~al.\/}(2010)Meliga, Sipp \& Chomaz]{Mel10}
{\sc Meliga, P., Sipp, D. \& Chomaz, J.-M.} 2010 Open-loop control of
  compressible afterbody flows using adjoint methods. {\em Physics of Fluids\/}
  {\bf 22}~(5), 054109.

\bibitem[Monkewitz(1988)]{mon88bluff}
{\sc Monkewitz, P.~A.} 1988 A note on vortex shedding from axisymmetric bluff
  bodies. {\em Journal of Fluid Mechanics\/} {\bf 192}, 561--575.

\bibitem[Nishioka \& Sato(1978)]{nis78}
{\sc Nishioka, M. \& Sato, H.} 1978 Mechanism of determination of the shedding
  frequency of vortices behind a cylinder at low reynolds numbers. {\em Journal
  of Fluid Mechanics\/} {\bf 89}, 49--60.

\bibitem[Passaggia {\em et~al.\/}(2012)Passaggia, Leweke \& Ehrenstein]{pas12}
{\sc Passaggia, P.-Y., Leweke, T. \& Ehrenstein, U.} 2012 Transverse
  instability and low-frequency flapping in incompressible separated boundary
  layer flows: an experimental study. {\em Journal of Fluid Mechanics\/} {\bf
  703}, 363--373.

\bibitem[Sipp(2012)]{Sip12}
{\sc Sipp, D.} 2012 Open-loop control of cavity oscillations with harmonic
  forcings. {\em Journal of Fluid Mechanics\/} {\bf 708}, 439--468.

\bibitem[Sipp \& Lebedev(2007)]{sip07}
{\sc Sipp, D. \& Lebedev, A.} 2007 Global stability of base and mean flows: a
  general approach and its applications to cylinder and open cavity flows. {\em
  Journal of Fluid Mechanics\/} {\bf 593}, 333--358.

\bibitem[Strykowski \& Sreenivasan(1990)]{Stry90}
{\sc Strykowski, P.~J. \& Sreenivasan, K.~R.} 1990 On the formation and
  suppression of vortex `shedding' at low {R}eynolds numbers. {\em Journal of
  Fluid Mechanics\/} {\bf 218}, 71--107.

\bibitem[Taneda(1956)]{Tan56}
{\sc Taneda, S.} 1956 Experimental investigation of the wakes behind cylinders
  and plates at low {R}eynolds numbers. {\em Journal of the Physical Society of
  Japan\/} {\bf 11}~(3), 302--307.

\bibitem[Thiria \& Wesfreid(2007)]{thi07}
{\sc Thiria, B. \& Wesfreid, J.~E.} 2007 Stability properties of forced wakes.
  {\em Journal of Fluid Mechanics\/} {\bf 579}, 137--161.

\bibitem[Tritton(1959)]{tri59}
{\sc Tritton, D.~J.} 1959 Experiments on the flow past a circular cylinder at
  low reynolds numbers. {\em Journal of Fluid Mechanics\/} {\bf 6}, 547--567.

\bibitem[Verma \& Mittal(2011)]{ver11}
{\sc Verma, A. \& Mittal, S.} 2011 A new unstable mode in the wake of a
  circular cylinder. {\em Physics of Fluids\/} {\bf 23}~(12), 121701.

\bibitem[Zielinska {\em et~al.\/}(1997)Zielinska, Goujon-Durand, Du\v{s}ek \&
  Wesfreid]{zie97}
{\sc Zielinska, B. J.~A., Goujon-Durand, S., Du\v{s}ek, J. \& Wesfreid, J.~E.}
  1997 Strongly nonlinear effect in unstable wakes. {\em Phys. Rev. Lett.\/}
  {\bf 79}, 3893--3896.

\end{thebibliography}

\end{document}